\documentclass[aps,pra,reprint,nofootinbib,twocolumn]{revtex4-2}
\usepackage{amsmath,amsthm,amssymb,mathrsfs}
\usepackage{physics}
\usepackage{bbold}
\usepackage{url}
\usepackage[dvipsnames,table]{xcolor}
  \definecolor{purple}{RGB}{128,0,128}
  \definecolor{codegreen}{rgb}{0,0.6,0}
  \definecolor{codegray}{rgb}{0.5,0.5,0.5}
  \definecolor{codepurple}{rgb}{0.58,0,0.82}
  \definecolor{backcolour}{rgb}{0.95,0.95,0.92}
  \definecolor{purple}{RGB}{128,0,128}
\usepackage[caption=false]{subfig}
\usepackage{listings}
  
  \lstdefinestyle{mystyle}{
    language=Python,
    backgroundcolor=\color{backcolour},   
    commentstyle=\color{codegreen},
    keywordstyle=\color{blue},
    numberstyle=\tiny\color{codegray},
    stringstyle=\color{codepurple},
    basicstyle=\ttfamily\footnotesize,
    breakatwhitespace=false,
    breaklines=true,
    captionpos=b,
    keepspaces=true,
    numbersep=5pt,
    showspaces=false,
    showstringspaces=false,
    showtabs=false,
    tabsize=2,
    escapechar=\%,
    frame=single
}
\usepackage{graphicx, wrapfig} 
\graphicspath{ {images/} }
\usepackage[hidelinks]{hyperref}

\newcommand{\id}{\mathbb{1}}
\newcommand{\ii}{\mathrm{i}}

\begin{document}

\title{A versatile neural-network toolbox for testing Bell locality in networks}

\author{Antoine Girardin}
\affiliation{Department of Applied Physics, University of Geneva, 1211 Geneva, Switzerland}

\author{Mohammad Massi Rashidi}
\affiliation{Department of Applied Physics, University of Geneva, 1211 Geneva, Switzerland}

\author{Géraldine Haack}
\affiliation{Department of Applied Physics, University of Geneva, 1211 Geneva, Switzerland}

\author{Nicolas Brunner}
\affiliation{Department of Applied Physics, University of Geneva, 1211 Geneva, Switzerland}

\author{Alejandro Pozas-Kerstjens}
\affiliation{Department of Applied Physics, University of Geneva, 1211 Geneva, Switzerland}

\begin{abstract}
Determining whether an observed distribution of events generated in a quantum network is Bell local, i.e., if it admits an alternative realization in terms of independent local variables, is extremely challenging.
Building upon [\href{http://dx.doi.org/10.1038/s41534-020-00305-x}{npj Quantum Inf. 6, 70 (2020)}], we develop a software solution that parameterizes local models in networks via neural networks.
This allows one to leverage optimization tools available from the machine learning community in the search of network Bell nonlocality.
Our solution applies to arbitrary networks, is easy to use, and includes technical improvements that significantly increase performance compared to previous implementations.
We apply it to investigate nonlocality in several networks hitherto unexplored, providing insights on the corresponding quantum nonlocal sets and suggesting concrete, promising realizations of quantum nonlocal correlations.
\end{abstract}

\maketitle

\section{Introduction}

Identifying situations in which classical and quantum physics predict different results (or, in modern words, finding a \textit{quantum advantage}) was once a task reserved exclusively for researchers interested in the foundations of physics \cite{Bell64}.
However, with the advent of quantum technologies, this work becomes increasingly relevant for experimental demonstrations and practical applications.

An unambiguous demonstration of quantum advantage is the observation of Bell nonlocality \cite{Bell64,Brunner2014}.
This concept is relevant in situations with multiple, non-communicating parties.
Bell nonlocality has led to applications, notably within the device-independent framework \cite{Acin2007,colbeck,Pironio2010}.
Recently, and with the practical aim of adapting these results to the experimental scenarios expected to be available in the future, Bell nonlocality has been investigated in the setting of networks.
In these setups, several independent sources distribute systems to different subsets of the parties, which can perform joint measurements on all the systems they receive \cite{Branciard2010,Fritz_2012,Tavakoli_Review}.

In Bell-type experiments, the distant parties perform local measurements on their quantum subsystems. The resulting statistics are the main object of interest.
In particular, when this data cannot be reproduced by a local model (in which all the sources distribute classical random variables instead of quantum systems), then it is said to demonstrate Bell nonlocality.
Although the characterization of local models in multipartite scenarios with a single source distributing shares of a single multipartite system to the parties is well understood (see, e.g., Ref.~\cite{Brunner2014}), this is not the case in networks.
There, the presence of independent sources makes a full characterization extremely challenging \cite{Tavakoli_Review}.
For this reason, the search for explicit local models has been limited to brute-force methods \cite{branciard2012bilocal,da_Silva_2023,da_Silva_2025}.

Given the complexity of the problem, a promising approach was proposed in Ref.~\cite{Krivachy_nn_2020}, which parameterized local models via neural networks, thereby leveraging the computational toolboxes developed in the area of machine learning.
This approach has turned to be quite successful, notably by helping to identify candidate realizations for network nonlocality \cite{Abiuso2022,Boreiri2023,Polino2023,wang2024,Boreiri2025,Krivachy2025} that in some cases have led to subsequent formal proofs \cite{Pozas_2023}.
However, the investigation in Ref.~\cite{Krivachy_nn_2020} and the associated code implementations were limited to a specific network, composed of three parties and three bipartite sources, coined the triangle network (see Fig.~\ref{fig:scheme_networks}).

In the present work, we build upon the approach of Ref.~\cite{Krivachy_nn_2020} and develop a generic software solution for constructing local models in networks using neural networks.
This solution can in principle handle arbitrary networks, and features additional improvements over previous software, leading to significantly enhanced performance in terms of accuracy and running time.
We illustrate its power by showing improvements over the results obtained in Ref.~\cite{Krivachy_nn_2020} and, most importantly, we use it to search for local models in four- and five-partite networks that were previously out of reach.
In these scenarios, and also in the triangle network, we identify promising candidates of quantum Bell nonlocal correlations. We conclude with a discussion and open questions.

\section{Scenario and problem}
\begin{figure*}[t]
    \null\hfill
    \subfloat[]{
        \centering
        \includegraphics[width=0.19\linewidth]{"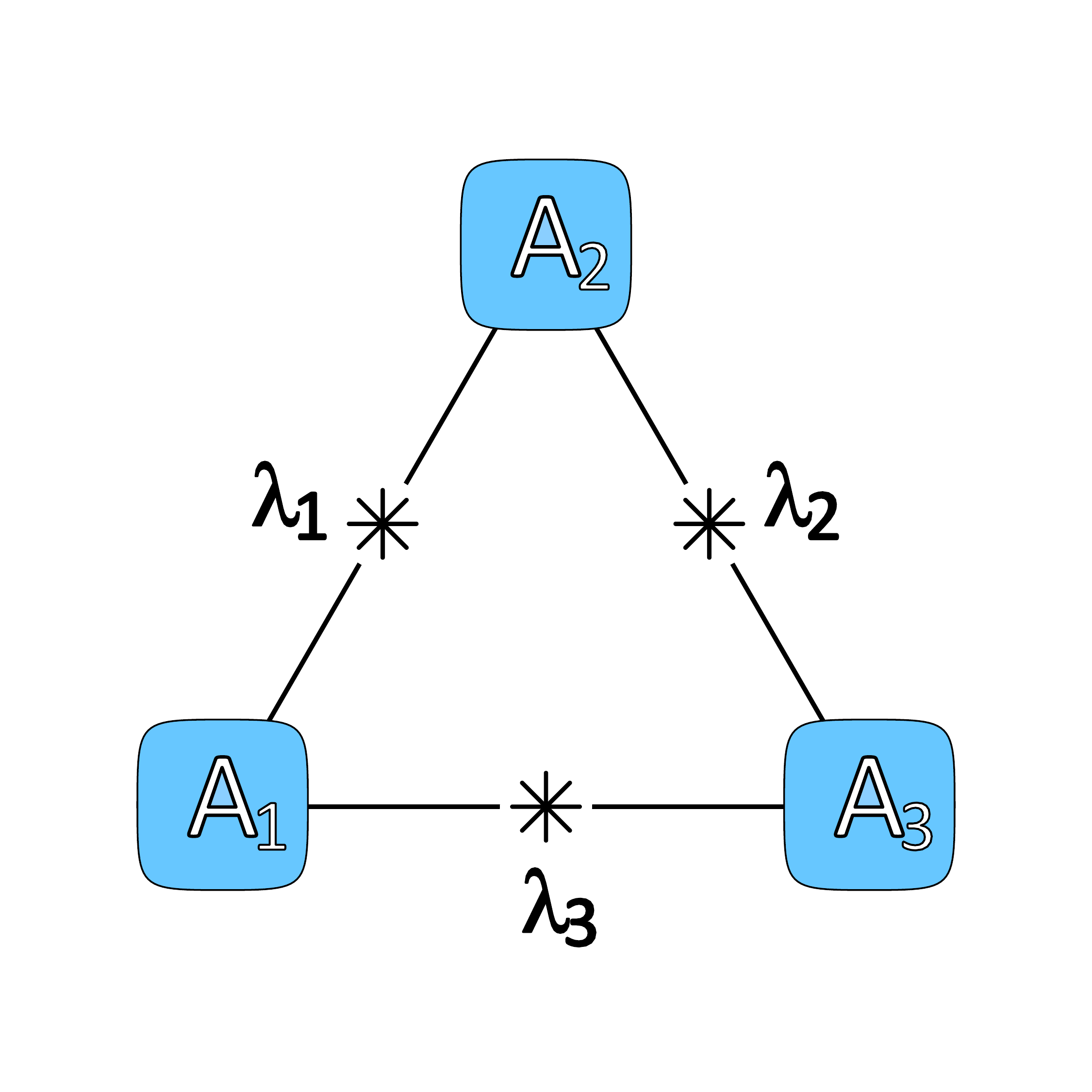"}
        \label{fig:scheme_networks:triangle}
    }
    \hfill
    \subfloat[]{
        \centering
        \includegraphics[width=0.19\linewidth]{"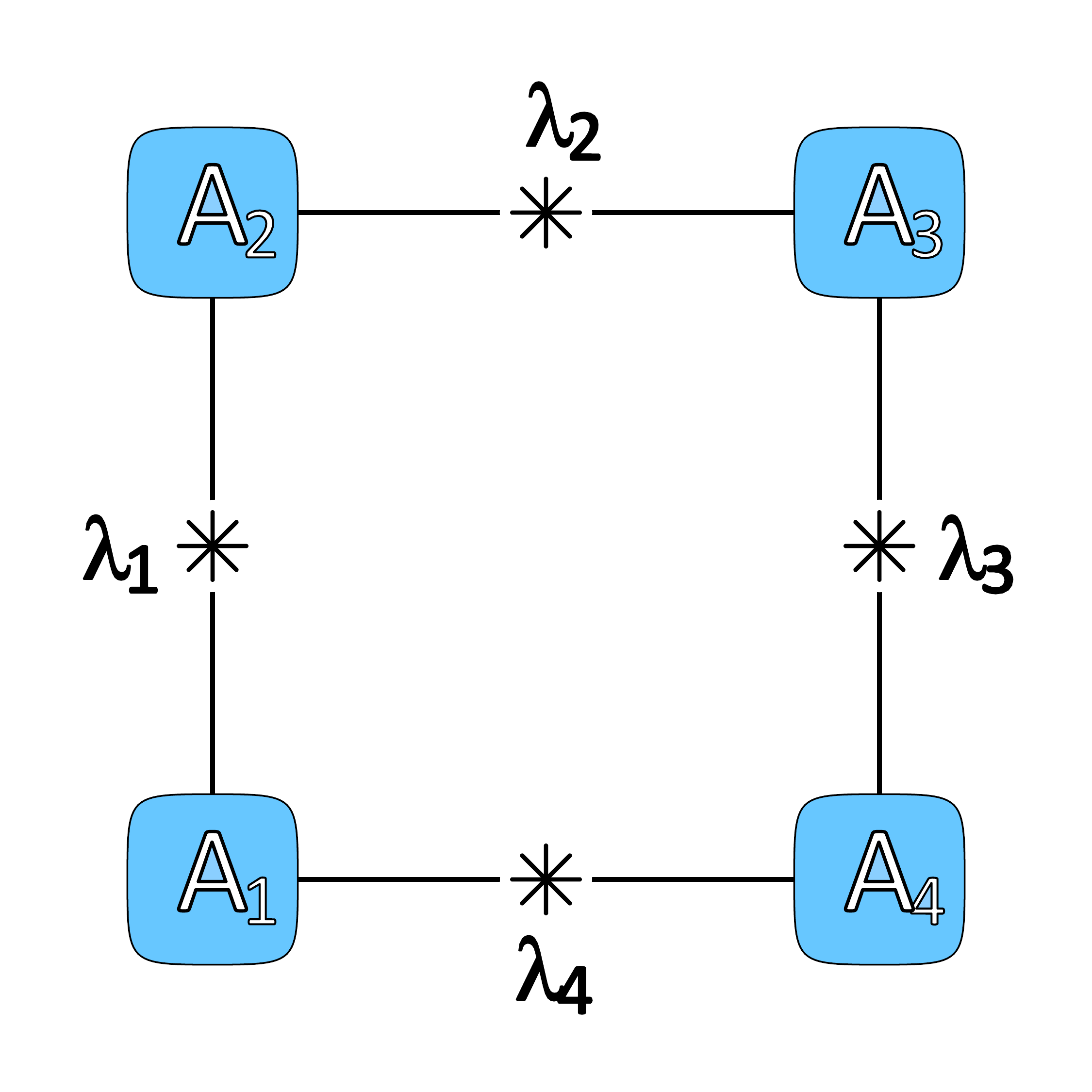"}
        \label{fig:scheme_networks:square}
    }
    \hfill
    \subfloat[]{
        \centering
        \includegraphics[width=0.19\linewidth]{"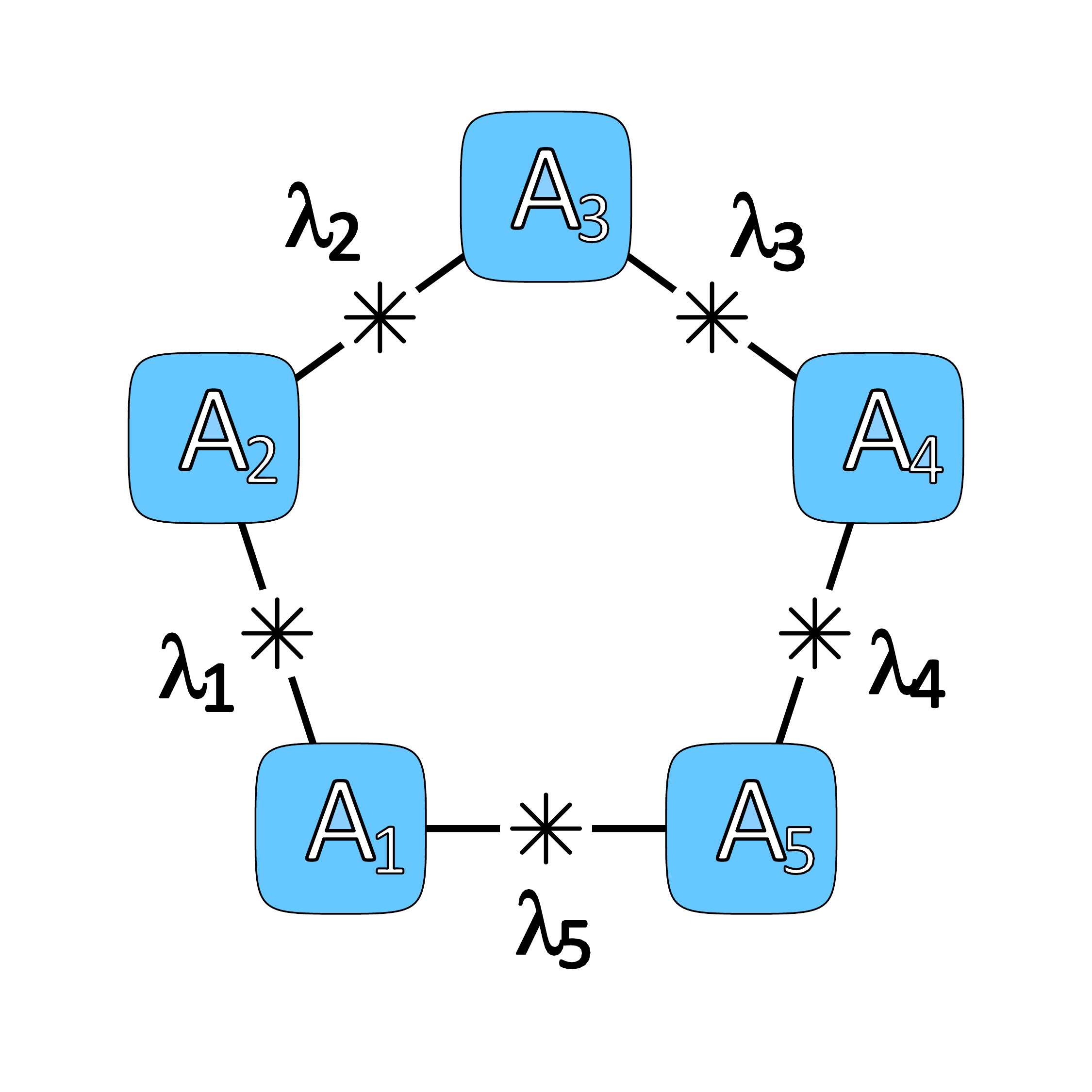"}
        \label{fig:scheme_networks:pentagon}
    }
    \hfill
    \subfloat[]{
        \centering
        \includegraphics[width=0.19\linewidth]{"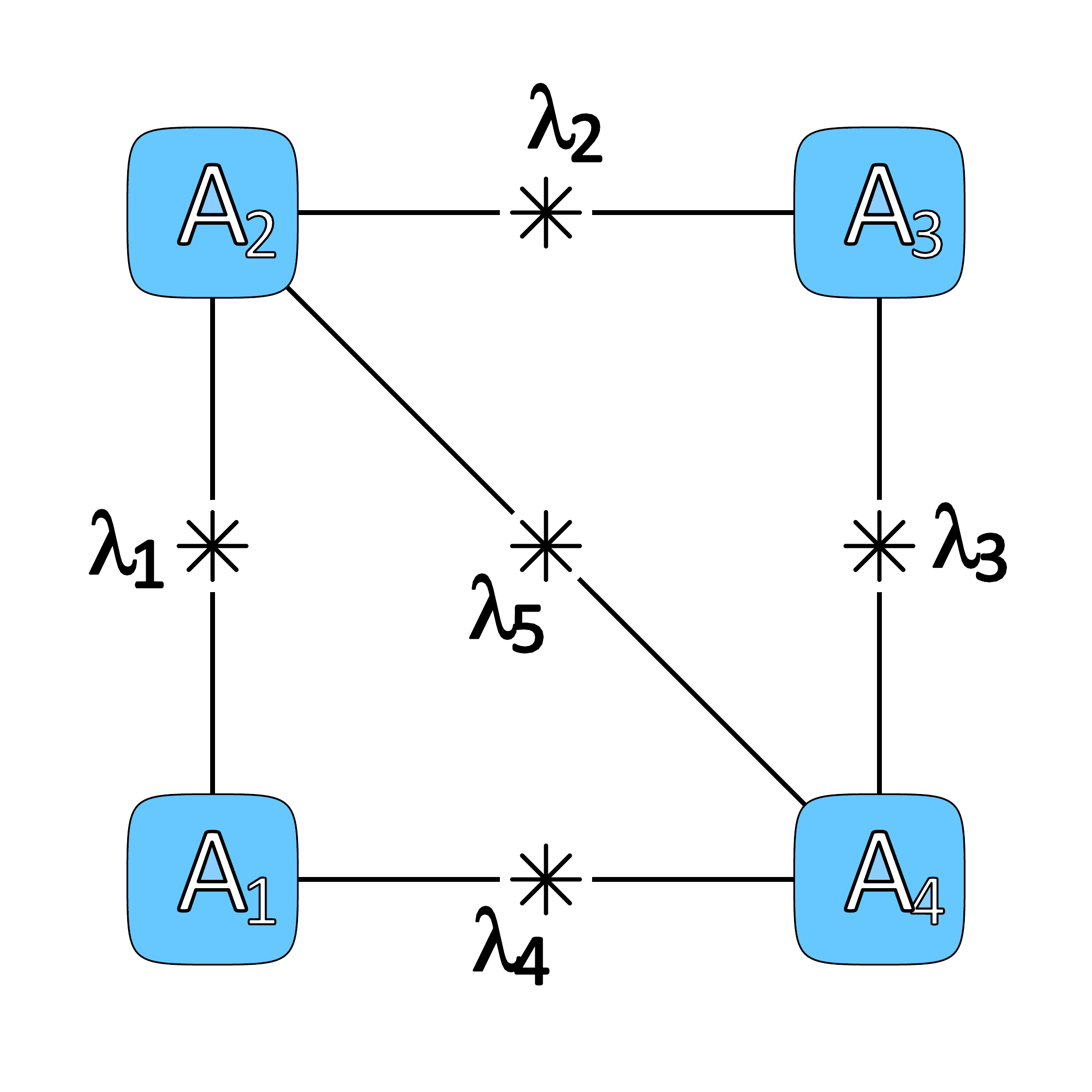"}
        \label{fig:scheme_networks:square1diag}
    }
    \hfill\null
    \caption{Scheme of the different networks we investigate in this work: \protect\subref{fig:scheme_networks:triangle} the triangle, \protect\subref{fig:scheme_networks:square} the square, \protect\subref{fig:scheme_networks:pentagon} the pentagon, \protect\subref{fig:scheme_networks:square1diag} the square with a diagonal.
    The small stars represent sources that distribute systems to the parties (blue squares), which process them to produce their corresponding outcomes.
    In this work we investigate whether the distributions of outcomes admit local models, which are of the form given in Eq.~\eqref{eq:network_local}.
    }
    \label{fig:scheme_networks}
\end{figure*}

We consider networks with $n$ parties, denoted $A_i$, connected by $m$ sources $S_j$ (see Fig.~\ref{fig:scheme_networks} for some simple examples).
Each source distributes a physical system (a shared random variable or a quantum state) to a subset of the parties.
Each party produces an output, denoted $a_i\in\{1,\dots,o_i\}$, based on the different systems it receives.
The main object of interest is the joint distribution over possible outputs, $p(a_1,\dots,a_{n})$. 

In this work, we will focus on distributions $p(a_1,\dots,a_{n})$ arising from quantum realizations.
This is, we will study distributions generated in networks where each source $S_j$ distributes a quantum state, $\ket{\psi_j}$, while each party $A_i$ performs a joint measurement on all the subsystems it receives, given by a set of operators $ \{M_{a_i} \}_{a_i=1}^{o_i}$ such that $M_{a_i}\succeq0$ and $\sum_{a_i} M_{a_i} = \openone$.
The resulting output distribution is given by the Born rule, i.e.,
\begin{equation}
    p(a_1,\dots,a_{n}) = \bra{\Psi}  \bigotimes_{i=1}^n  M_{a_i} \ket{\Psi},
    \label{eq:born}
\end{equation}
where $\ket{\Psi} = \bigotimes_{j=1}^m \ket{\psi_j}$.
Note that, in practice, one needs to carefully permute the subsystems in the above equation in order to carefully match the sources and the measurements.

The question we ask is whether a given quantum distribution of the form of Eq.~\eqref{eq:born} admits a local model, i.e., whether it can be realized also in the case when the sources distribute systems that are described by classical physics instead.
These realizations, called \textit{local models}, are defined as follows: let each source $S_j$ produce a random variable $\lambda_j$.
Crucially, we assume all sources to be independent from each other, as in the above quantum realizations, hence the local variables $\lambda_j$ are also considered to be independent.
This is the defining difference between local models in networks and in multipartite Bell scenarios, and the source of the difficulty of searching for local models in networks \cite{Branciard2010,Tavakoli_Review}.
Let us denote $S_{A_i}\subseteq \cal{S}$ is the subset of sources connected to the party $A_i$, where $\mathcal{S}=\{\lambda_1,\dots,\lambda_m\}$ is the set of all local variables.
We say that an observed distribution $p(a_1,\dots,a_{n})$ admits a local model if it can be expressed in the following way:
\begin{equation}\label{eq:network_local}
    p(a_1,\dots,a_{n}) = \int\prod_{j=1}^{m} d\lambda_{j} p_j(\lambda_j)\prod_{i=1}^{n}p_{A_i}(a_i|S_{A_i}) \,.
\end{equation}
Here $p_j(\lambda_j)$ denotes the probability distribution of the local variable $\lambda_j$ and $p_{A_i}(a_i|S_{A_i})$ represents the response function of party $A_i$ that produces output $a_i$ given the received local variables $S_{A_i}$. 

Given a distribution $p(a_1,\dots,a_{n})$ obtained, e.g., from a quantum model, finding out whether it admits a decomposition of the form \eqref{eq:network_local} is typically very challenging. Formally, the complexity of the problem stems from the independence of the sources in Eq.~\eqref{eq:network_local}, i.e., from the fact that the distribution of the local variables must factorize.
This can be understood in geometrical terms: we seek to characterize the set of distributions admitting a decomposition of the form of \eqref{eq:network_local}, which is a nonconvex set.
In contrast, the situation is much simpler in ``standard'' Bell scenarios, which feature a single source connecting all parties; in this case, the problem is linear and the set of Bell-local distributions is convex, so it can be efficiently characterized \cite{Brunner2014}.

Several methods have been developed for testing the existence of local models for distributions in networks, such as the inflation method \cite{wolfe2019inflation,boghiu2022}, non-linear Bell inequalities \cite{branciard2012bilocal,Chaves2016,Rosset2016} or entropic inequalities \cite{Chaves2012,Weilenmann_2018}, as well as methods based on self-testing \cite{Renou_2019,renou2022network,Sekatski2023,Boreiri_2025}.
Yet, in many cases of interest, these techniques remain ineffective for detecting nonlocality.
Hence, one needs to resort to numerical techniques.
For the simplest networks, a brute-force search over local models is in principle possible \cite{da_Silva_2023,da_Silva_2025}, since the cardinality of the local variables can be upper bounded without loss of generality \cite{rosset2017universal}.
Such search is, however, out of reach beyond the simplest cases.
Alternatively, one can resort to methods based on neural networks \cite{Krivachy_nn_2020}, as we will investigate in this paper.

Before moving on, let us point that our presentation here focuses on networks in which each party performs a single, fixed measurement.
This is in contrast with, for instance, the bipartite Bell scenario in which the parties need to perform at least two measurements each in order to reveal nonlocality \cite{Bell64}.
Remarkably, multiple measurements per party are not needed to demonstrate nonlocality in networks \cite{Fritz_2012,Pozas2023b,Boreiri2023,Sekatski2023}.
Thus, this type of networks has been considered with great attention, and is known as ``correlation scenarios'' in the literature \cite{Fritz_2012,navascues2017inflation}.
It is, of course, important to also consider networks with inputs.
Our approach can be readily adapted to such scenarios, since it is always possible to map a network with inputs to another where the inputs are replaced by one additional source that distributes the input and one additional party that announces it, as described in Ref.~\cite{Fritz_2012}.

\section{Software solution}
In this work we parameterize local models in networks via neural networks, building upon the approach developed in Ref.~\cite{Krivachy_nn_2020}.
This parameterization allows one to address in an efficient manner the problem discussed in the previous section: given a target distribution, construct a local distribution as close as possible to it. 
We develop a software solution providing a number of improvements over the implementation of Ref.~\cite{Krivachy_nn_2020}, namely
\begin{itemize}
    \item The code is general and user-friendly: it can be readily applied to any network topology, and contains a simple interface as well as a module for computing distributions from a given quantum network with ease.
    \item We improve the performance and runtime due to an active adaptation of the number of samples during the training. This makes training typically faster and more efficient.
    \item We include auxiliary tools for, e.g., easily computing the resulting distribution from the states distributed and the measurements performed, or visualizing the response functions learned.
\end{itemize}

Below, we start by presenting the approach, and then discuss in more detail each improvement of our implementation.

\begin{figure*}[t]
    \null\hfill
    \subfloat[]{
        \includegraphics[width=0.35\linewidth]{"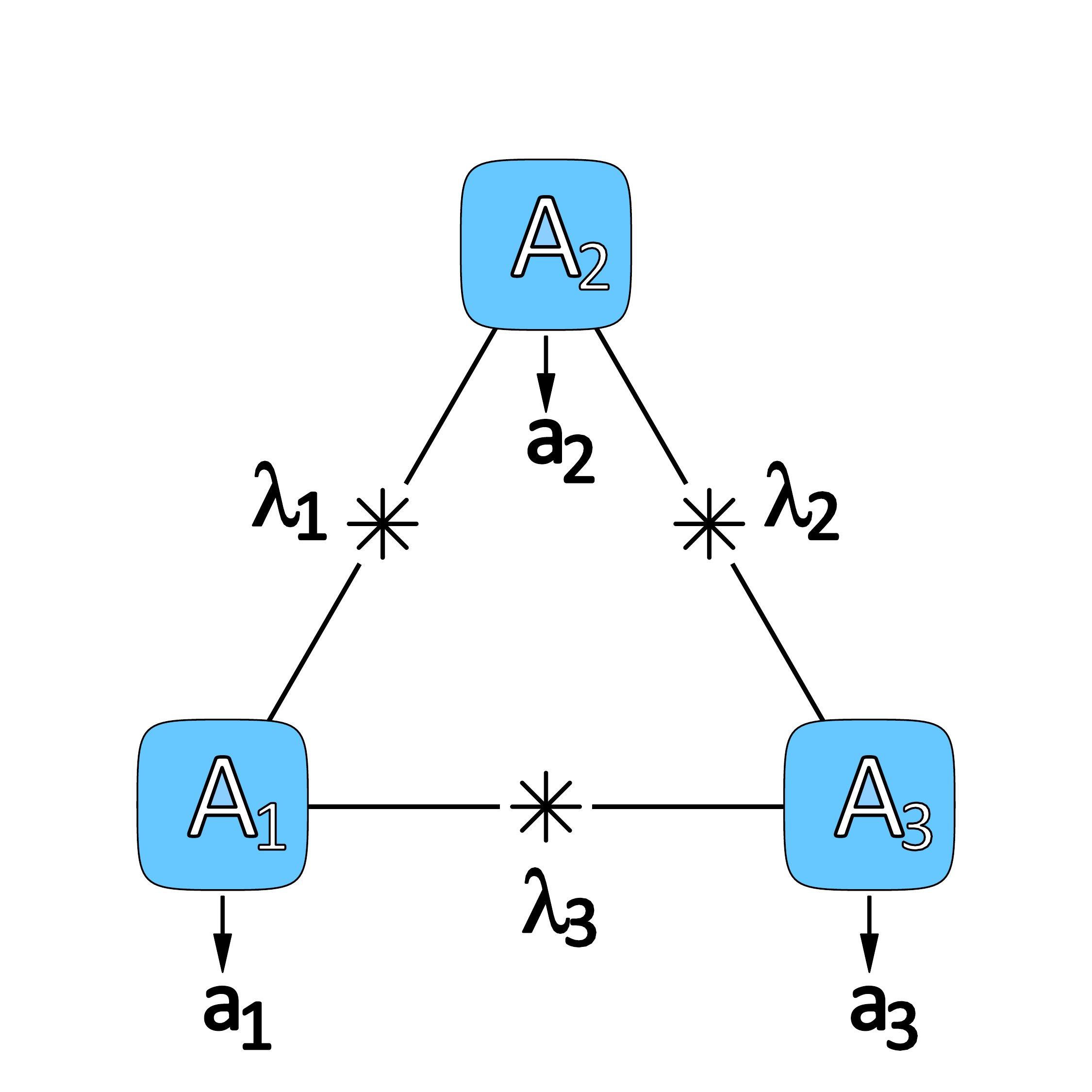"}
        \label{fig:triangle2}
    }
    \hfill
    \subfloat[]{
        \includegraphics[width=0.45\linewidth]{"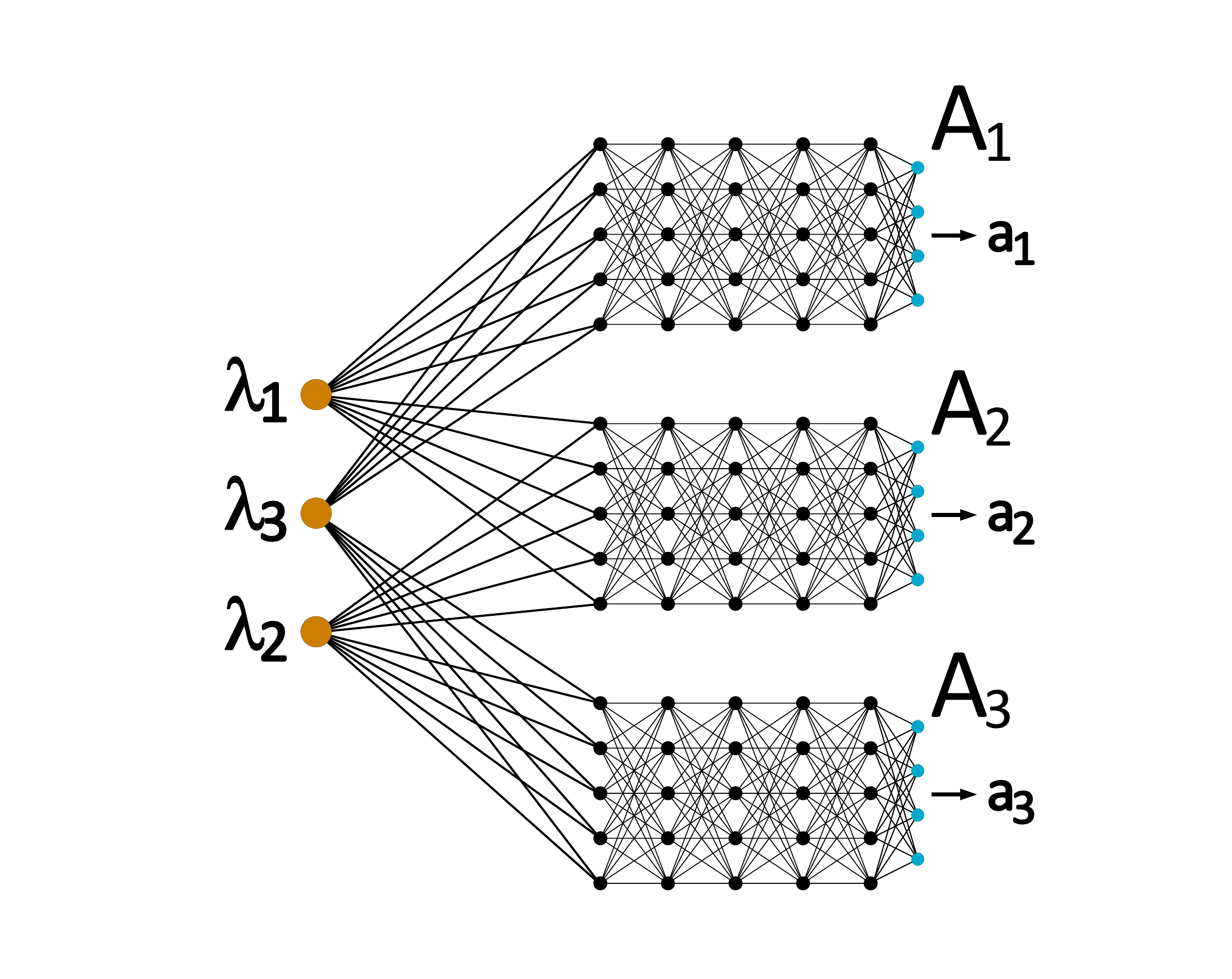"}
        \label{fig:trianglenet}
    }
    \hfill\null
    \caption{Illustration of the neural-network ansatzes. Panel \protect\subref{fig:triangle2} depicts the triangle network. In panel \protect\subref{fig:trianglenet} we show a neural network that reproduces the topology of the triangle: the input layer has three neurons (one per source), and then three blocks of feed-forward architecture capture the response functions of each of the parties. The topology of the network is captured by the fact that the information in the input neurons is sent, not to all, but only to the corresponding parties.
    }
    \label{fig:triangle}
\end{figure*}

\subsection{Neural network ansatzes for local models}
The core idea of the approach is to parameterize Eq.~\eqref{eq:network_local} by a neural network that intrinsically captures the topology of the physical network, thereby accommodating the independent sources in a natural manner.
Motivated by this, the ansatz begins with a layer of $m$ neurons, each of which represents one of the $m$ sources in the network, and will be allowed to take values in a continuous interval.
In traditional feedforward neural networks, these neurons will be connected to subsequent layers in an all-to-all manner.
Instead, we connect each set of neurons representing the sets of sources $S_{A_i}$ to a different processing block, that consists of a small feedforward neural network that produces the vector of probabilities $\{p_{A_i}(a_i|S_{A_i})\}_{a_i=1}^{o_i}$ for party $A_i$ given the values of the hidden variables that are connected to it in the network.
An illustration of a possible ansatz for the triangle network is given in Fig.~\ref{fig:triangle}.
It is well known that increasing the number of hidden layers (we will refer to this number as \textit{depth}), as well as the number of neurons in each of the layers (the \textit{width}) of each block allows for a larger family of possible response functions, but is also computationally more costly.

Contrasting this construction with Eq.~\eqref{eq:network_local}, it is immediately noticeable that (i) at this stage there is no reference to the distribution of the local hidden variables (which, moreover, we will consider fixed, see below), and (ii) the local hidden variables are continuous, while it is known that in Eq.~\eqref{eq:network_local} they can be taken to be discrete without loss of generality \cite{rosset2017universal}.
The class of probability distributions created when setting some fixed distributions for the hidden variables, nevertheless, encompasses all distributions of the form of Eq.~\eqref{eq:network_local} in the limit of arbitrarily large processing blocks.
It is easy to see that any probability distribution over a discrete set of events can be produced from a probability distribution over a continuous interval by suitably discretizing the interval.
Moreover, this processing can be performed locally at each party, and thus be absorbed into the processing block representing $p_{A_i}$. 
Therefore, we can safely use the neural network ansatzes to describe distributions that admit network-local models.

The neural network stores the local model implicitly.
In order to access the resulting distribution, it is necessary to sample the local hidden variables and pass them through the response functions.
This produces an empirical reconstruction of the distribution,
\begin{equation}
    \hat{p}_\textsc{nn}(a_1,\dots,a_n)=\frac{1}{N_s}\sum_{k=1}^{N_s} \prod_{i=1}^np_{A_i}(a_i|S_{A_i,k})
    \label{eq:estimate}
\end{equation}
(where $S_{A_i,k}$ denotes the sampled values of the local hidden variables in $S_{A_i}$ in step $k$), that converges to the actual distribution, $p_\textsc{nn}$, in the limit $N_s\rightarrow\infty$.
This empirical approximation depends on the parameters of the neural network, and can be used to define a suitable notion of distance to a target distribution, $p_\textsc{t}$ ---coming, e.g., from a quantum realization via Eq.~\eqref{eq:born}---, that will be minimized by (variants of) gradient descent.
Thus, the precision with which $\hat{p}_\textsc{nn}$ approximates $p_\textsc{nn}$, controlled by the number of samples $N$, is an important factor that can be exploited to improve the performance and efficiency of the gradient descent optimization.
One of the main improvements of our software package with respect to previous implementations is a careful tuning of $N_s$ during the fitting process.
We provide details on this feature in Section \ref{sec:adaptive}.

To obtain the results of this manuscript we use two different forms for the loss functions to be minimized.
The first is the Kullback–Leibler (KL) divergence, defined as
\begin{equation}\label{eq:KL_div}
    d_{KL}(p_\textsc{t}||\,\hat{p}_\textsc{nn})=\sum_i p_\textsc{t}(i) \log\left(\frac{p_\textsc{t}(i)}{\hat{p}_\textsc{nn}(i)}\right).
\end{equation}
This quantity is not (strictly speaking) a distance because it is not symmetric.
Yet, it is useful here, since it quantifies the loss of information when replacing the true target distribution by the approximate one reconstructed from the neural network.
Alternatively, we also use the Euclidean distance between the two distributions, i.e.
\begin{equation}\label{eq:eucl_dist}
    d_{eucl}(p_\textsc{t},\hat{p}_\textsc{nn})=\sqrt{\sum_i |p_\textsc{t}(i)-\hat{p}_\textsc{nn}(i)|^2}.
\end{equation}
We will use both metrics to present the results, but we observe empirically that the KL divergence typically achieves better performance for training the neural network.
For this reason, we always initially train the neural networks using the KL divergence as loss.
For the cases where we display the Euclidean distance, we optimize this loss for a number of rounds in a second stage after an initial training that optimizes the KL divergence.

The final goal is to make a statement about the target distribution, in particular whether it admits a local model or not.
In general, it is not possible to recover the target distribution exactly via sampling, so our method cannot give a formal proof.
However, if the neural network manages to construct a local model at a small distance to the target distribution, we are guaranteed that there exists a local model for a distribution at least that close to the target.
This being said, in certain cases, the local strategy is simple enough to be constructed analytically from the neural network.

In the case that the neural network is not able to find a distribution close to the target distribution, two possibilities arise: (i) the target distribution is indeed network nonlocal, or (ii) the neural network failed at finding the underlying local model.
While one cannot distinguish between these two cases, this information is nevertheless useful, and can serve to guide future research, in particular towards deriving a formal proof using, e.g., inflation methods (as in, e.g., Ref.~\cite{Pozas_2023}).
Interestingly, but perhaps unsurprisingly, when the optimization fails to find a distribution close to the target, the optimal distribution reconstructed by the algorithm strongly depends on the choice of loss function. 

In case the target distribution is suspected to be nonlocal, a further useful procedure consists in investigating how the distance between the target and the reconstructed distributions evolves when noise is added to the target distribution.
For example, when considering a target distribution built from a quantum model, one typically adds noise to the states emitted by the quantum sources, e.g., starting from pure entangled states and adding white noise.
If the initial target distribution (for the pure states) is indeed nonlocal, one expects to see, as a function of the amount of noise inserted, first a decrease in the distance between the target distribution and the optimal fit, followed by a flat segment corresponding to a small and approximately constant distance when the target distribution admits a local model; see, e.g., Refs.~\cite{Krivachy_nn_2020,Abiuso2022} and below for examples.
We observe that the shape of these curves depends on which distance is used.
For the Euclidean distance, the decay is typically linear, with a sharp kink when entering the local region.
For the KL divergence, the transition is smoother, yet there is often a greater contrast between local and nonlocal distributions.

\subsection{Interface}\label{sec:contributions}
The software solution that we develop is available at \cite{repo}.
Its main part is the definition and optimization of the neural-network local models.
This is implemented using the PyTorch platform~\cite{pytorch}, which provides a user-friendly interface, as well as enabling the leverage of all the machine learning toolbox for optimizing the free parameters of the model.

An illustration of a simple example of use is depicted in Code snippet~\ref{fig:code}.
We consider the triangle network.
Following Fig. \ref{fig:triangle}, we denote the parties $A_1$, $A_2$ and $A_3$, their outputs $a_1$, $a_2$ and $a_3$, and the sources $\lambda_1$, $\lambda_2$ and $\lambda_3$.
The network topology is specified in the \texttt{config} dictionary, whose keys are strings denoting the parties, and the corresponding values are tuples of two objects.
The first object is a list of strings, which denote the sources that feeds to the corresponding party.
In Code snippet 1, note that each variable appears in two different parties, hence enforcing the structure of the triangle network.
The second one is an integer, that denotes the number of outcomes for the corresponding party.
In Code snippet 1, this is four outcomes for each party.
The target distribution is then specified as a vector with all the probabilies $p_\textsc{t}(a_1,\dots,a_n)$ in row-major order.\footnote{This is the default order in, e.g., the \texttt{flatten} function in NumPy.}
The model is an instance of the \texttt{NeuralNetwork} class that contains all the necessary information to build the neural network, such as the width and depth of the response functions or the loss function (all the parameters can be found in the documentation in Ref.~\cite{repo}).
Finally, the function \texttt{train\_model\_and\_save} starts the training of the model and saves it into a file after finishing.

\lstset{numbers=left}
\begin{lstlisting}[language=Python, float, caption=Minimal illustration of how to use the code in the case of the triangle network with four outputs per party.,label={fig:code}]
from module.nn_sampling.build_nn import NeuralNetwork, train_model_and_save

# Define network as a dictionary with entries
# party: (list_of_sources, n_outcomes)
config = {
    "a1": (["lambda1", "lambda3"], 4),
    "a2": (["lambda2", "lambda1"], 4),
    "a3": (["lambda3", "lambda2"], 4)
}

# The target distribution as a list
target_distribution = [...] 

# Create neural network
model = NeuralNetwork(config, target_distribution)

# Train
train_model_and_save(model)
\end{lstlisting}

Many additional options can be used to control the shape of the neural network and the training.
We provide examples on how to use them in Ref.~\cite{repo}.

\subsection{Optimizing the sampling}\label{sec:adaptive}
The optimized neural networks store an implicit definition of the local response functions of the parties, so the only way to access the distribution is via its estimation from samples.
This process is subject to errors coming from finite statistics.
Increasing the number of samples will reduce this sampling error, but will also slow down the training process.
It is therefore important to choose carefully the amount of samples used for the estimation.
Our software adapts the number of samples during the fitting, in order to make the latter more efficient and accurate.
In this section we describe the reasoning behind the heuristics that we use for this purpose.

Consider a target distribution, $p_\textsc{t}$, and the distribution encoded by a neural network, $p_{\textsc{nn}}$.
The goal is to minimize a proxy for the difference between these two distributions, for example a distance, which we denote in general by $\norm{p_\textsc{t} -p_\textsc{nn}}$.
In practice, one does not have access to $p_{\textsc{nn}}$, but rather to an empirical estimate of it via Eq.~\eqref{eq:estimate}.
Let us call this estimate $\hat{p}_\textsc{nn}$.
For any distance functional admitting the triangle inequality (such as the Euclidean distance), we can see that
\begin{equation}
    \begin{aligned}
        \norm{p_\textsc{t} - p_\textsc{nn}} &=\, \norm{p_\textsc{t} - \hat{p}_\textsc{nn} + \hat{p}_\textsc{nn} - p_\textsc{nn}} \\
        &\leq\, \norm{p_\textsc{t} - \hat{p}_\textsc{nn}} + \norm{\hat{p}_\textsc{nn} - p_\textsc{nn}}.
    \end{aligned}
\end{equation}

In the right-hand side of the equation above, the first term, $\norm{p_\textsc{t} - \hat{p}_\textsc{nn}}=\mathcal{L}$ is, in fact, the function we optimize during the fitting, while the second term, $\norm{\hat{p}_\textsc{nn} - p_\textsc{nn}}=d_s$, quantifies the error due to finite statistics.
In order to have meaningful gradients during the fitting process, we want to use an amount of samples such that the dominant term is $\mathcal{L}$.
Thus, after obtaining some value of the loss function $\mathcal{L}^{(i)}$ in an iteration, we will set the number of samples for the next round so that the expected sampling error, $d_s^{(i+1)}$, is smaller than $\mathcal{L}^{(i)}$.

Then, what remains is to find a relation between $d_s$ and the number of samples, $N_s$.
We find empirically that the sampling error decreases with the number of samples as $d_s=1/\sqrt{N_s}$ for the Euclidean distance, and as $d_s=N_o/(2N_s)$ for the KL divergence, where $N_o$ is the number of possible outcomes (see Appendix~\ref{app:sampling_error}).
These scalings (albeit not the concrete prefactors that we find) are theoretically justified.
First, from the central limit theorem, one can expect the sampling error to decrease as $1/\sqrt{N_{s}}$.
Moreover, for the Euclidean distance, it can be shown that $d_s \leq \frac{1}{\sqrt{N_s}} \sqrt{ 1-\frac{1}{N_o}} $, where $N_o$ denotes the number of outputs.\footnote{This relation can be shown by realizing that one can model the amount of times that a given set of outcomes $(a_1,\dots,a_n)$ is seen when sampling as a binomial distribution with probability of success $p(a_1,\dots,a_n)$. Calling this amount of times $N_{a_1,\dots,a_n}$, we have that
\begin{equation*}
    \begin{aligned}
        \norm{p_\textsc{t} - \hat{p}_\textsc{nn}}_2^2 &= \frac{1}{N_s^2}\sum_{a_1,\dots,a_n}|N_s p(a_1,\dots,a_n)-N_{a_1,\dots,a_n}|^2 \\
        &= \frac{1}{N_s^2}\sum_{a_1,\dots,a_n}\text{Var}(N_{a_1,\dots,a_n})\\
        &=\frac{1}{N_s}\sum_{a_1,\dots,a_n}p(a_1,\dots,a_n)\left[1-p(a_1,\dots,a_n)\right].
    \end{aligned}
\end{equation*}
This quantity is upper bounded by the uniform distribution, $p(a_1,\dots,a_n)=N_o^{-1}\,\,\,\forall\,a_1,\dots,a_n$. Substituting, one arrives to $d_s \leq \frac{1}{\sqrt{N_s}} \sqrt{ 1-\frac{1}{N_o}}$.
We believe that this relation is well known, but failed to find a reference describing it.}
The KL divergence is not a metric, and thus, it does not admit a triangle inequality.
Yet the scaling of the sampling error that we observe is in agreement with known results (see e.g. Ref.~\cite[Eq. 6]{Mardia2019}).

Note also that these scalings can be useful to determine the potential (non-)locality of a target distribution: If, when increasing the number of samples, the distance between the target distribution and the empirical one obtained from the neural network converges to zero with the scalings given, it means that Eq.~\eqref{eq:estimate} is dominated by finite-sampling effects up to the maximum number of samples used, and thus the neural network is likely capturing the target distribution. If, on the contrary, one observes convergence to a non-zero value, or a behavior different from the scalings above, then one can ascertain that the network has not captured the target. Note that this can happen either due to the target distribution being nonlocal, or by artifacts such as the neural network not being expressive enough or the optimization not reaching the global optimum of the loss function.

In practice, we want to have the sampling error to be smaller than the new loss, but we do not want to take an excessive amount of samples.
In order to control this, we define a bias parameter, $B$, such that we choose the number of samples at iteration $i+1$ to be $N_s^{(i+1)}=(B/\mathcal{L}^{(i)})^2$ when minimizing the Euclidean distance and $N_s^{(i+1)}=B N_o/\mathcal{L}^{(i)}$ for the KL divergence.
Empirically, we observe that values of $B$ between 2 and 4 perform well.
For all the results presented in Sections \ref{sec:benchmark} and \ref{sec:larger}, we use an initial value of $B=4$, that we increase in one unit after a number of iterations without improvement in the optimization until a maximum value of $B=10$.

\subsection{Additional utilities}
Our software solution also features a number of utilities that ease the preparation of data and the interpretation of results.

First, it includes a small module for the computation of probability distributions in the required format by applying Eq.~\eqref{eq:born} to a given set of states and measurements.
The relevant function in this module is \texttt{quantum\_network\_fct}, which produces the distribution from the states and measurements.
Importantly, the Hilbert spaces of the states and POVMs must be carefully matched.
This matching is specified by the user via the argument \texttt{order\_hs\_sources}, that consists of a list, $\ell$, with the integers $[0,1,\dots,T-1]$, with $T=\sum_{i=1}^N|S_{A_i}|$ the total amount of Hilbert spaces (i.e., particles) in the network.
This list is ordered such that particle $\ell_l$ is matched with the $l$-th Hilbert space for the POVMs.
For instance, for the triangle network of Fig.~\ref{fig:triangle}, if we provide the list of states as $[\lambda_{1},\lambda_{2},\lambda_{3}]$ and the list of POVMS as $[A_1,A_2,A_3]$, the list provided as \texttt{order\_hs\_sources} should be $[5,0,1,2,3,4]$.

Second, for the case of parties that are connected to two sources, it includes a function, \texttt{plot\_strats}, that plots the implicit strategies encoded by the neural network as a function of the values of the hidden variables.
This is particularly useful, since in certain situations it allows to easily infer analytic expressions for the strategies (see, e.g., \cite[Fig. 3]{Krivachy_nn_2020}).

\section{Benchmarking and improved accuracy}\label{sec:benchmark}

\begin{figure*}[t]
    \null\hfill
    \subfloat[]{
        \centering
        \includegraphics[width=0.45\linewidth]{"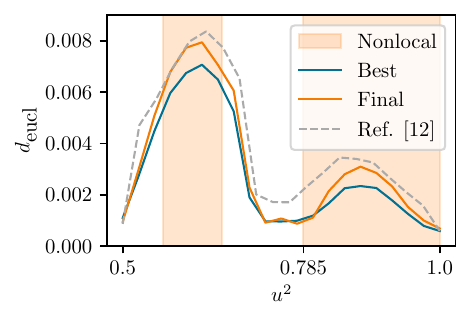"}
        \label{fig:results_rgb4:u}
    }
    \hfill
    \subfloat[]{
        \centering
        \includegraphics[width=0.336\linewidth]{"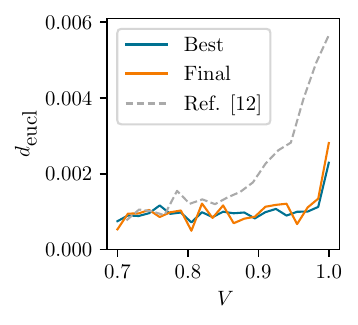"}
        \label{fig:results_rgb4:vis}
    }
    \hfill\null
    \caption{
        Results of the neural network for the RGB4 family of distributions.
        Panel \protect\subref{fig:results_rgb4:u} shows the distributions obtained by distributing $\ket{\psi_+}$ states in all sources and performing the RGB4 measurement \eqref{eq:TCmeasurements} in all parties of the triangle network, as a function of the parameter $u^2$ in the measurements.
        Panel \protect\subref{fig:results_rgb4:vis} depicts the distances for the distributions with measurements fixed to $u^2 = 0.85$, and the sources distributing Werner states of visibility $V$.
        The light gray curve is a reproduction of Fig. 5 in Ref.~\cite{Krivachy_nn_2020}.
        The orange curves depict the final distances obtained with our software, while the blue curves show the best distances obtained during the training (which may be affected by sampling errors).
        All results are obtained using a depth of 4 and a width per party of 60.
        Training is performed for a maximum of $10^4$ iterations, stopping earlier if there has been no improvement over $10^3$ iterations.
        }
    \label{fig:results_rgb4}
\end{figure*}

In order to benchmark our algorithm, we start by investigating a specific class of quantum distributions, some of which are known to be nonlocal \cite{Renou_2019}.
The nonlocality of these distributions, and in particular their robustness to noise, have also been discussed in Ref.~\cite{Krivachy_nn_2020}.
Here we demonstrate that our algorithm provides more accurate results than those reported in Ref.~\cite{Krivachy_nn_2020}, showing that the noise robustness of the distributions is, in fact, much weaker than previously thought.

Specifically, we consider the family of distributions on the triangle network introduced in Ref.~\cite{Renou_2019}, which from now on we term RGB4.
The nonlocality of such distributions has been investigated in a series of works \cite{renou2022network,Pozas_2023,Sekatski2023,Boreiri2023,Boreiri_2025}.
The quantum model is the following: Each source produces the two-qubit Bell state $\ket{\psi_+} = (\ket{01}+\ket{10})/\sqrt{2}$.
Each party performs a projective measurement, with the following eigenvectors: 
\begin{equation}
    \{ \ket{00} ;  u \ket{01}+v\ket{10} ;  v \ket{01}-u\ket{10} ;  \ket{11}  \}
    \label{eq:TCmeasurements}
\end{equation}
with a real parameter $u \in [0,1]$ and $v= \sqrt{1-u^2}$.
This distribution is proven to be nonlocal for $u^2 \in ] 0.785,1[$ \cite{renou2019limits}, and $u \in ]0.563,0.657[$ \cite{Pozas_2023}.
For $u=1$ and $u^2=0.785$, the distribution admits a local model \cite{Renou_2019}.

Scanning the parameter over the interval $u \in [1/\sqrt{2},1]$, we run our algorithm to find the closest local distribution. The results are shown in Fig.~\ref{fig:results_rgb4:u}, which depicts the minimal euclidean distance as a function of $u^2$.
We observe results fully consistent with Ref.~\cite{Krivachy_nn_2020}, with some improvement in terms on the distance.
However, the most important aspect of these results is that they are obtained with an out-of-the-box configuration (the neural network for each response function has a width of 60 neurons and a depth of 4 layers, and the KL divergence is minimized using the Adam optimizer), while the results of Ref.~\cite{Krivachy_nn_2020} required a careful use of different optimizers and loss functions in different stages of the training \cite{private}.

Next, we investigate the noise robustness of the nonlocal distribution obtained with $u^2=0.85$ (the most nonlocal/distant point in the right interval in Fig.~\ref{fig:results_rgb4:u}).
This is, now we make each source to distribute a Werner state of the form
\begin{equation}
    \rho = V \ket{\psi_+}\bra{\psi_+} + (1-V) \openone/4 \, ,
\end{equation}
where $V$ denotes the visibility.
In Fig.~\ref{fig:results_rgb4:vis}, we plot the distance as a function of the visibility.
Starting from $V=1$ (no noise), we expect to see first a monotonous decrease of the distance, followed by a plateau when the distribution becomes local.
The point where the behaviour changes provides an indication about the critical visibility $V^*$, the largest visibility at which the distribution admits a local model.
Here, we observe significant improvements over the results of Ref.~\cite{Krivachy_nn_2020}.
The latter suggested a critical visibility around $V^* \simeq 0.89$, while our results indicate a much higher critical visibility, $V^* \simeq 0.99 $, much closer to the bound proven in Ref.~\cite{Boreiri_2025}, $V^* <0.9945$.
Again, these results are obtained with an out-of-the-box configuration.
This indicates that the adaptive sampling significantly improves the estimation, and with it the accuracy of the fit via gradient descent.

\section{Exploration of larger networks}\label{sec:larger}
We now use our algorithm to investigate nonlocality beyond known examples, in particular in larger networks, which so far have remained unexplored.
Initially, we consider ring networks with three, four and five parties (see Figs.~\ref{fig:scheme_networks:triangle}-\ref{fig:scheme_networks:pentagon}).
We parametrize quantum models and look for the distributions that appear to have the most nonlocality, in the sense of having the largest distance to a local distribution.
These models involve bipartite states and bipartite measurements.
Later, we move to a more sophisticated network (shown in Fig.~\ref{fig:scheme_networks:square1diag}), which is a square with an additional source connecting two opposite parties.
Here, the quantum model still features bipartite sources, but two parties now perform tripartite measurements.

Before presenting the results, let us introduce the quantum models that we use.
For the bipartite states, we consider two-qubit maximally entangled states of the form
\begin{equation}\label{eq:rotated_state_1}
    \ket{\psi^{\theta}}=(e^{\ii\frac{\theta}{2}X}\otimes \id) \ket{\phi_+} = \cos\frac{\theta}{2}\ket{\phi_+} +\ii\sin\frac{\theta}{2}\ket{\psi_+} \,.
\end{equation}
with $\theta \in [0,\pi]$.
This allows interpolating between two orthogonal Bell states, namely $\ket{\phi_+}=(\ket{00}+\ket{11})/\sqrt{2}$ when $\theta=0$ and $\ket{\psi_+}$ for $\theta=\pi$.

For bipartite measurements, we use the family of joint measurements with tetrahedral symmetry defined in Ref.~\cite{Tavakoli_2021}.
Specifically, the four measurement eigenstates are given by
\begin{equation}\label{eq:eBSM}
    \ket{\Phi_b^{\mu}}= \frac{\sqrt{3}+e^{\ii\mu}}{2\sqrt{2}}\ket{\vec{m}_b, -\vec{m}_b} + \frac{\sqrt{3}-e^{\ii\mu}}{2\sqrt{2}}\ket{-\vec{m}_b, \vec{m}_b}   \, ,
\end{equation}
with $b=0,\dots,3$.
Here, $\ket{ \pm \vec{m}_b}$ are states associated to the Bloch vectors forming a regular tetrahedron on the Bloch sphere; see \cite{Tavakoli_2021} for details.
This allows one to interpolate between the ``elegant joint measurement'' \cite{gisin2019entanglement}, which is achieved at $\mu=0$, and (a local rotation of) the usual Bell basis at $\mu=\pi/2$.

When exploring non-locality in the network of Fig.~\ref{fig:scheme_networks:square1diag}, we will also need to define three-qubit measurements.
We will consider two different classes of them.
First, the generalization of the Bell state measurement to three qubits, namely a measurement in a basis of  GHZ states, defined by the eigenstates
\begin{equation}\label{eq:GHZ_meas}
    M_{i_1,i_2,i_3} = X^{i_1}\otimes X^{i_2}\otimes Z^{i_3} \left[\frac{1}{\sqrt{2}}(\ket{000}+\ket{111})\right],
\end{equation}
where $X=\ket{0}\bra{1}+\ket{1}\bra{0}$ and $Z=\ket{0}\bra{0}-\ket{1}\bra{1}$ are the well-known Pauli matrices.
The second class generalises the ``token counting'' measurement, used e.g. in the RGB4 distribution and given in Eq. \eqref{eq:TCmeasurements}.
The eigenstates of the measurement are given by
\begin{equation}\label{eq:tc_3qb}
    \begin{aligned}
        M_0 = \ket{000},
        &\quad \begin{pmatrix} M_1 \\ M_2 \\ M_3 \end{pmatrix} = \frac{1}{\sqrt{3}}\begin{pmatrix} 1 & 1 & 1 \\ 1 & \omega & \omega^2 \\ 1 & \omega^2 & \omega \end{pmatrix} \begin{pmatrix} \ket{001} \\ \ket{010} \\ \ket{100} \end{pmatrix}, \\
        M_{4+i} &= X\otimes X\otimes X \cdot M_i,
    \end{aligned}
\end{equation}
with $\omega = e^{\ii\frac{2\pi}{3}}$.
This measurement can be seen as a generalization of the bipartite \eqref{eq:TCmeasurements} for $u=1/\sqrt{2}$.

\begin{figure*}[t]
    \centering
    \subfloat[]{
        \includegraphics[width=0.48\linewidth]{"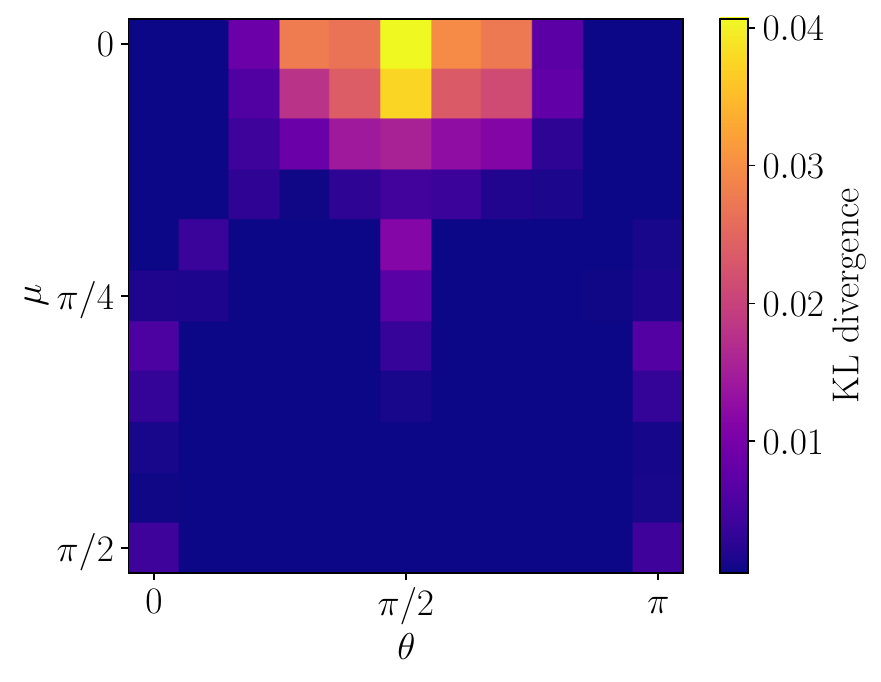"}
        \label{fig:results_2d_plots:triangle}
    }
    \subfloat[]{
        \includegraphics[width=0.48\linewidth]{"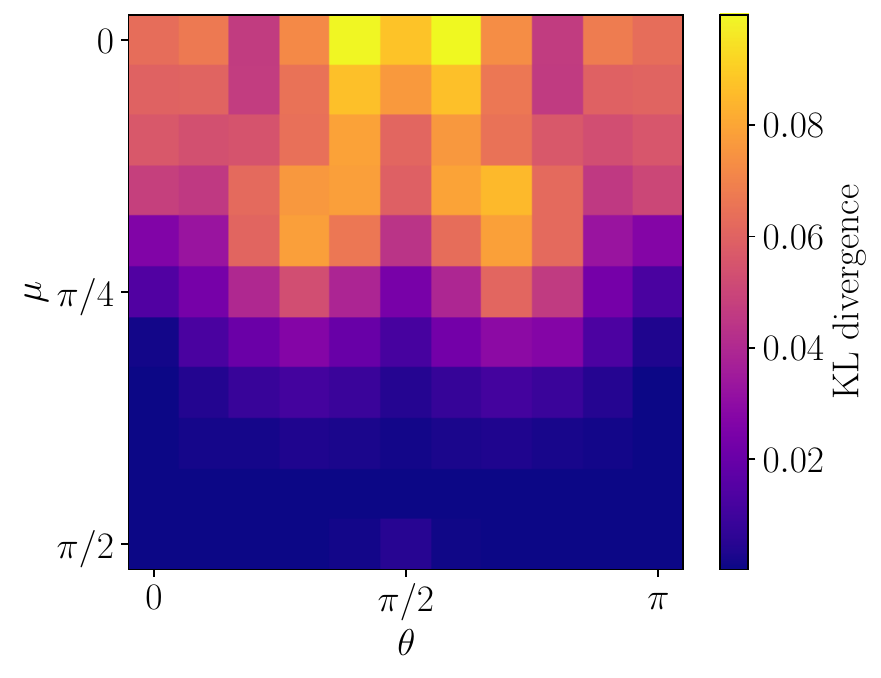"}
        \label{fig:results_2d_plots:square}
    }
    \\
    \subfloat[]{
        \includegraphics[width=0.48\linewidth]{"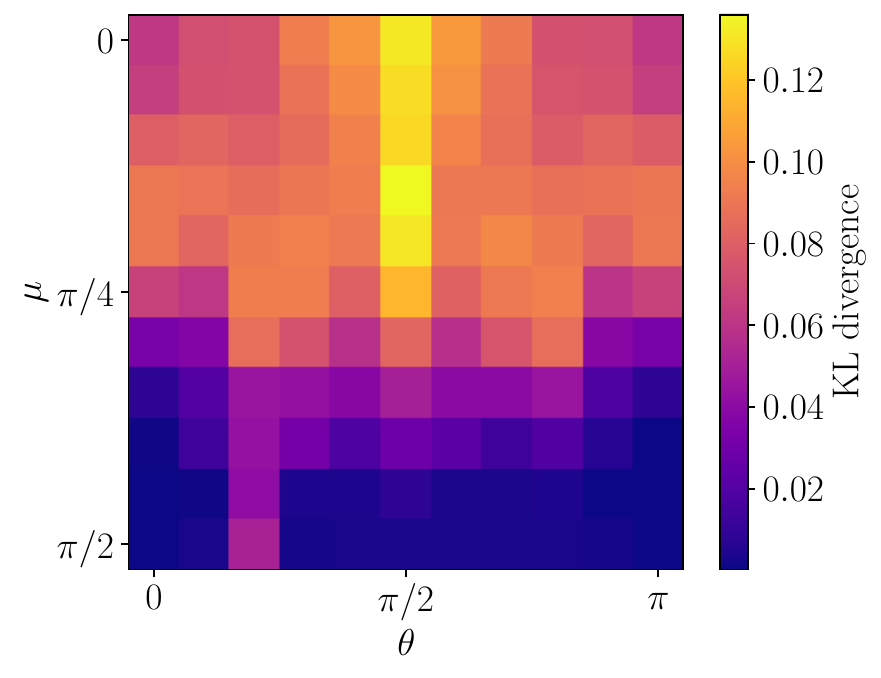"}
        \label{fig:results_2d_plots:pentagon}
    }
    \subfloat[]{
        \includegraphics[width=0.48\linewidth]{"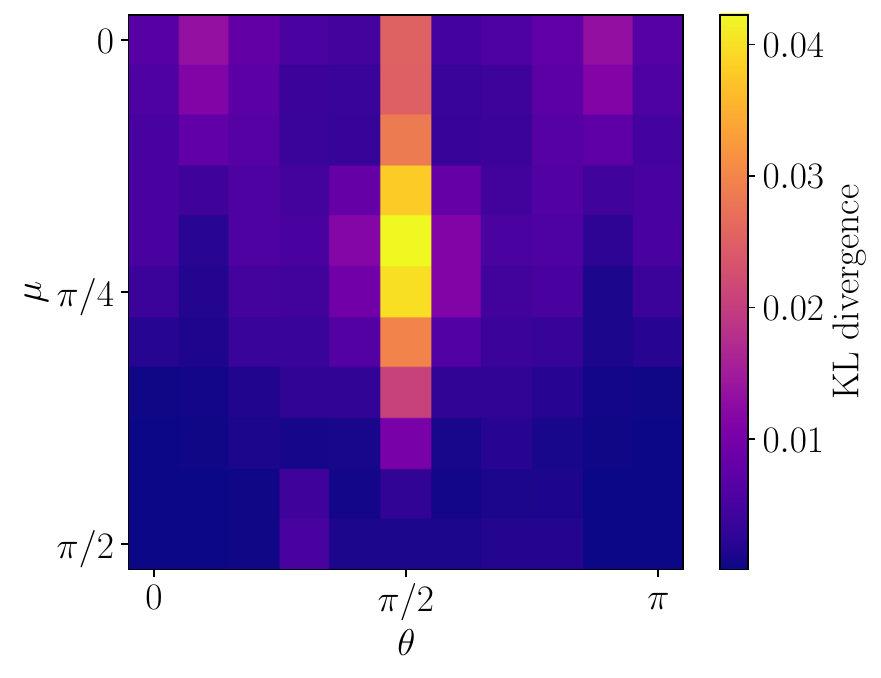"}
        \label{fig:results_2d_plots:pentagon3}
    }
    \caption{
        Results of the neural network when scanning the family of measurements given in Eq.~\eqref{eq:eBSM} and the family of states in Eq.~\eqref{eq:rotated_state_1} in \protect\subref{fig:results_2d_plots:triangle} the triangle network, \protect\subref{fig:results_2d_plots:square} the square network, \protect\subref{fig:results_2d_plots:pentagon} the pentagon network, and \protect\subref{fig:results_2d_plots:pentagon3} the pentagon with three outputs.
        All results are obtained using a depth of 4 and a width per party of 60.
        Training is performed for a maximum of $10^4$ iterations, stopping earlier if there has been no improvement over $10^3$ iterations.
    }
    \label{fig:results_2d_plots}
\end{figure*}

For all the results reported below, we use neural network models of depth 4 and width 60.
Heuristically we observed that slightly varying the depth (2-4-6) and/or the width (40-60-80) consistently produced worse results.
However, this does not necessarily mean that the model size that we used is optimal.
In order to have strong evidence for optimality, one should perform a complete hyperparameter tuning, including not just model width and depth but also, e.g., initialization, optimizer, and scheduling.
While this in-depth study lies outside the scope of this work, we note that the fact that the library is implemented in PyTorch means that one can use well-established hyperparameter tuning tools, such as Optuna \cite{optuna} and Ray Tune \cite{raytune}, to perform it.

\subsection{Ring networks}
We begin by considering quantum models built from the above families of states and measurements for ring networks from the triangle to pentagon.
Since both the states and the measurements are characterized by a single parameter each, we perform a full scan of both, minimizing the KL divergence.
The corresponding results are presented as 2D plots in Fig.~\ref{fig:results_2d_plots}.
Darker regions indicate distributions that are close to a local one (hence likely to admit a local model) while regions in lighter colours indicate the possible presence of nonlocality.
Here, we use the KL divergence as a metric to enhance the contrast between the types of behaviors.
Panel \subref{fig:results_2d_plots:triangle} depicts the results for the triangle network, panel \subref{fig:results_2d_plots:square} for the square, and panel \subref{fig:results_2d_plots:pentagon} for the pentagon.
Note that in all these cases we use four-outcome measurements, so we have distributions with four outputs per party.
In panel \subref{fig:results_2d_plots:pentagon3} we also depict results for the pentagon network, but in this case the parties coarse-grain the outcomes corresponding to $b=0$ and $b=1$ into a single outcome, so the target distributions have three outcomes per party.

Interestingly, the plots hint at a symmetry around $\theta=\frac{\pi}{2}$.
For fixed measurements, the distributions when using states with $\theta=\frac{\pi}{2}\pm\vartheta$ have the same probabilities, although in different supports that (at least in the case of the triangle) are not related to each other via outcome relabelings.
Thus, it is not immediately obvious that this apparent symmetry reveals some symmetry of the underlying distributions.

Starting with the triangle, i.e., panel \subref{fig:results_2d_plots:triangle}, we observe that the most promising candidate for nonlocality appears when using the elegant joint measurement ($\mu=0$) with the state for $\theta=\pi/2$.
The observed distance is in fact even slightly larger than for the ``elegant distribution'', where the EJM is combined with the Bell state $\ket{\psi_-}=(\ket{01}-\ket{10})/\sqrt{2}$, producing a non-local distribution \cite{Gitton2025}.
In Appendix~\ref{app:robustness} we perform a more thorough analysis of this distribution, studying its robustness to noise in the states.
Note that when the EJM is combined with $\ket{\phi_+}$ ($\theta=0$) or $\ket{\psi_+}$ ($\theta=\pi$), the resulting distribution is local.
Hence, we see that bipartite entanglement is not enough by itself since it is important to carefully match the states and measurements.
We tried proving the nonlocality of the distributions for $\mu=0,\,\pi/6,\,\pi/4,\,\pi/3$ using the optimized inflation codes of Ref.~\cite{Gitton2025} but, to the extent of our computing capabilities (which allow us to run the inflation consisting of two copies of two of the sources and three copies of the remaining one) we were not able to prove that these distributions were, in fact, nonlocal.
Also, we observe that the Bell state measurement ($\mu=\pi/2$) always leads to local distributions.
This is in accordance with the result of Ref.~\cite{Gatto_Lamas_2023}, which states distributions constructed from stabilizer states and measurements in the Clifford group always admit a local model.

Moving to the square network, depicted in panel \subref{fig:results_2d_plots:square}, we first observe that more combinations of states and measurements produce potentially nonlocal distributions.
This tendency is further confirmed when observing the results in the pentagon in panel \subref{fig:results_2d_plots:pentagon}.
We also see that the most likely nonlocal configuration seems to change.
For the square, combining the EJM with states just off of $\theta= \pi/2$ leads to the largest distance.
In the pentagon, the EJM still appears to lead to nonlocality, but even more nonlocality seems to be obtained for another measurement parameter (around $\theta=\pi/4$).
When coarse-graining the measurement outputs in panel \subref{fig:results_2d_plots:pentagon3}, we observe that the difference between the EJM and $\mu\approx\pi/4$ becomes stronger.
We also observe that the distance significantly increases with the number of parties, being three times larger for the pentagon than in the triangle.

Overall, we observe that the larger the network, the easier it seems to produce nonlocal distributions.
In the triangle, most distributions appear to be local (or, at least, very close to the local set), but when increasing the size of the network, most of the distributions appear to be nonlocal.
It would be interesting to verify if this conjecture is correct, or if it comes from a limitation of our procedure.
As the network size increases, the neural network that captures the corresponding model contains more parameters, and thus is more prone to fall in local minima during training. While the overall consistency of the different plots in Fig.~\ref{fig:results_2d_plots} is an indication against numerical imprecisions, it is not possible to rule this hypothesis out completely.
This could in principle be addressed by using larger neural networks and the subsequent increase in computational power.

In Appendix~\ref{app:morering} we present a similar analysis, but considering a different family of maximally entangled states, namely an interpolation between $\ket{\phi_-}=(\ket{00}-\ket{11})/\sqrt{2}$ and $\ket{\psi_-}$, while keeping the same measurements.
This leads to different distributions and results.

\subsection{Network with tripartite measurements}
We now look at the network illustrated in Fig.~\ref{fig:scheme_networks:square1diag}.
In this case, all sources are still bipartite, but two of the parties receive three systems, and thus have to perform tripartite measurements.
For this analysis we limit ourselves to a small selection of states and measurements in order to just illustrate the performance of the method.
Concretely, we consider only a handful of cases in which the sources distribute all the same state and the parties perform the same bipartite and tripartite measurements.
For the states we consider the four Bell states; for the bipartite measurements we consider the Bell state measurement and the measurement \eqref{eq:TCmeasurements} for $u^2=0.85$; and for the tripartite measurements we consider those given by Eqs.~\eqref{eq:GHZ_meas} and \eqref{eq:tc_3qb} as generalizations of the formers.

In particular, the distributions resulting from any of the Bell states and Bell and GHZ state measurements admit a local model \cite{Gatto_Lamas_2023}.
Thus, these distributions give us a baseline with respect to which we can assess whether a local realization is likely or not.
Comparing the KL divergence on a candidate realization with that corresponding to realizations that admit local models is another way to obtain indications of its potential nonlocality.

The results for all the states and measurements considered are shown in Table~\ref{tab:squarediag}.
Interestingly, in this part of the study we found that escaping from the local minimum given by the uniformly random distribution was more challenging and involved longer trainings.
This may be, in part, due to the increased complexity of the network and the associated neural-network models, and in part due to the large amount of zeros in the resulting distributions, which make the optimization of the KL divergence challenging since few empirical terms are used.
In order to escape this minimum, in this case the training for all distributions begins from a same model that is pre-trained on reproducing the distribution corresponding to $\ket{\phi_+}$ as the state, Eq.~\eqref{eq:TCmeasurements} for the bipartite measurements, and the GHZ measurement for the tripartite measurements, for $2\times10^3$ iterations.
Then, the training consists in ten consecutive sessions on the corresponding distribution, each of which is composed of three repetitions of $2\times10^3$ rounds of minimization of the KL divergence, $2\times10^3$ rounds of minimization of the Euclidean distance, and $2\times10^4$ new rounds of minimization of the KL divergence.

From the physical point of view, we observe several interesting phenomena.
First, starting from the local realizations (any Bell state, plus Bell and GHZ state measurements), it seems like changing only the tripartite measurements is not enough to produce potentially nonlocal distributions.
Second, if we change only the bipartite measurements instead, it is not clear that nonlocality can be produced, since the KL divergences obtained are consistently larger than those for local distributions, but the difference is not large enough to be a clear indication.
Finally, when all the measurements are inspired by token counting strategies (i.e., Eqs.~\eqref{eq:TCmeasurements} and \eqref{eq:tc_3qb}), the KL divergence is more than one order of magnitude larger than that corresponding to local distributions.
Thus, in this case, we conjecture that all the resulting distributions are nonlocal.
\begin{table}
    \centering
    \begin{tabular}{cc|c|c|c|c}
           &                                         & $\phi_+$ & $\phi_-$ & $\psi_+$ & $\psi_-$ \\ \hline
      BSM, & GHZ                                     & 0.0010 & 0.0017 & 0.0017 & 0.0011 \\
      BSM, & \eqref{eq:tc_3qb}                       & 0.0011 & 0.0012 & 0.0012 & 0.0014 \\
      \eqref{eq:TCmeasurements}, & GHZ               & 0.0020 & 0.0021 & 0.0022 & 0.0020 \\
      \eqref{eq:TCmeasurements}, & \eqref{eq:tc_3qb} & 0.0273 & 0.0562 & 0.0570 & 0.0219 \\
    \end{tabular}
    \caption{
        KL divergences between the neural network models and the distributions created in the network of Fig.~\ref{fig:scheme_networks:square1diag} using the states in the columns and the measurements in the rows. 
        Note that the configurations in the first row admit local models per Ref.~\cite{Gatto_Lamas_2023}, and thus serve as a baseline.
        All results are obtained using a depth of 4 and a width per party of 60.
        }
    \label{tab:squarediag}
\end{table}

\section{Discussion}
We have developed a software solution that allows for the systematic and fast search of local models for quantum realizations in networks.
Speed is achieved by parameterizing local models via neural networks, and making use of the tools (in particular, PyTorch) and the methods (for instance, momentum-based optimization or adaptive gradient updates) that are well-known in the machine learning community.

Building upon the concept and implementation of Ref.~\cite{Krivachy_nn_2020}, which focused exclusively on the triangle network, our solution accommodates any network structure, automatically building the parameterizations of the corresponding local models.
Moreover, we have implemented several additional features, such as adaptive sampling, which provide a significant improvement of the performance, both in terms of speed and accuracy.

It is important to keep in mind the limitations of these techniques.
First, the numerical nature of the optimization problem prevents the distance between the target and parametrized distributions to be exactly zero.
This means that our framework cannot directly lead to a formal proof of the existence of a local model.
From a more practical perspective, however, it is useful to show that a given target distribution is close enough to---and hence practically indistinguishable from---a local distribution.
One can become convinced of this closeness by, e.g., comparing the results with those obtained for distributions that are known to be local (as we have done in Fig.~\ref{fig:results_2d_plots} and Table~\ref{tab:squarediag}), or by observing the behavior of the distance between the empirical and target distributions when increasing the number of samples used to build the empirical distribution.
Second, failing to achieve a small distance between the target and the parametrized distributions cannot be considered a guarantee of nonlocality.
In particular, such a failure could be due to an insufficiently expressive neural network, or to getting stuck in local minima.
While having access to all the toolbox of PyTorch partly alleviates this issue, one can only gather mounting evidence of nonlocality.
In order to get a formal proof of nonlocality, one must eventually resort to alternative techniques. 

Despite these limitations, the software solution we develop allows for the exploration of large networks that were previously inaccessible, thereby identifying networks, states, and measurements that are promising for quantum advantage.
In particular, we have explored quantum nonlocality in more complex networks (some of them unexplored so far), featuring up to five parties and tripartite measurements.
We have identified several promising quantum distributions that appear to be strongly nonlocal in terms of noise robustness.
Importantly, our results indicate that maximally entangled states beyond the traditional Bell states seem to be an important and currently overlooked resource for network nonlocality.
Yet, this represents only numerical evidence, and the key step will be to provide analytical proofs of these results.
We have applied existing methods, in particular inflation \cite{wolfe2019inflation,Gitton2025}, but could not arrive at conclusive results.
We hope that future works will eventually provide such proofs.

Finally, it would also be interesting to investigate these questions for more sophisticated nonlocality tests.
Here we have focused on networks ``without inputs'', where each party performs a fixed measurement.
While our techniques are in principle directly amenable to networks with inputs (since any such network can be converted into a network without inputs by inserting additional sources, see Section 2), it may still be that inputs can be included more efficiently in practice.
Another interesting direction is nonlocality scenarios featuring extra classical communication:
Neural-network solutions have already been developed for investigating the classical communication cost of quantum correlations in the bipartite Bell scenario \cite{Sidajaya2023}.
Exploring more sophisticated networks with extra communication, such as the instrumental \cite{Chaves2018,VanHimbeeck2019} and Evans \cite{Lauand2025} networks, would therefore be interesting. 

\acknowledgments
We thank Tam\'as Kriv\'achy for helpful discussions and comments, and Fran\c{c}ois Fleuret for technical directions.
This work is supported by the Swiss National Science Foundation (grant numbers 192244 and 224561), and the Swiss State Secretariat for Education, Research and Innovation (SERI) under contract number UeM019-3. Computations were performed in part at the University of Geneva using the Baobab HPC service.

\bibliography{nnfornl.bib}

\onecolumngrid
\appendix
\setcounter{figure}{0}
\renewcommand{\thefigure}{\thesection\arabic{figure}}

\section{Empirical analysis of the sampling error}
\label{app:sampling_error}
One of the main improvements of our software is the automatic tuning of the number of samples taken to compute the empirical distribution in order to generate meaningful gradients for the optimizer.
As described in Section~\ref{sec:adaptive}, we choose the number of samples such that the expected error due to sampling is smaller than the loss obtained in the previous optimization step.
In order to choose this number, we need to know how the sampling error varies with the number of samples, both for the Euclidean distance and the Kullback-Leibler (KL) divergence, and taking into account the number of outcomes of the distribution, $N_o$.

\begin{figure*}[h]
	\includegraphics[width=0.45\linewidth]{"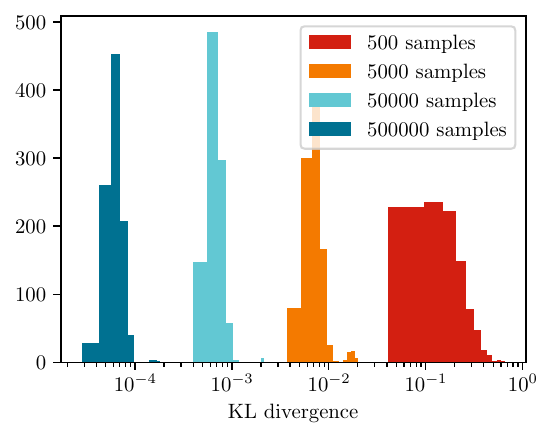"}
    \includegraphics[width=0.45\linewidth]{"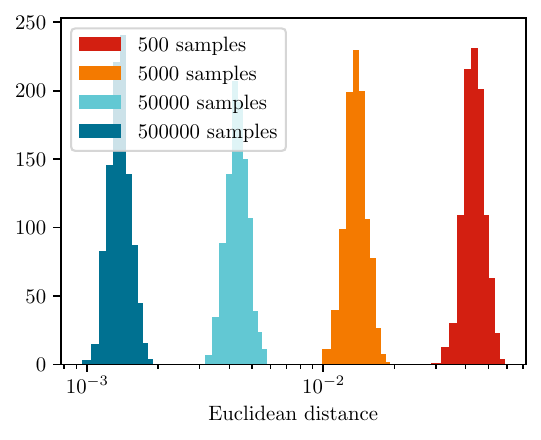"}
    \\
	\includegraphics[width=0.45\linewidth]{"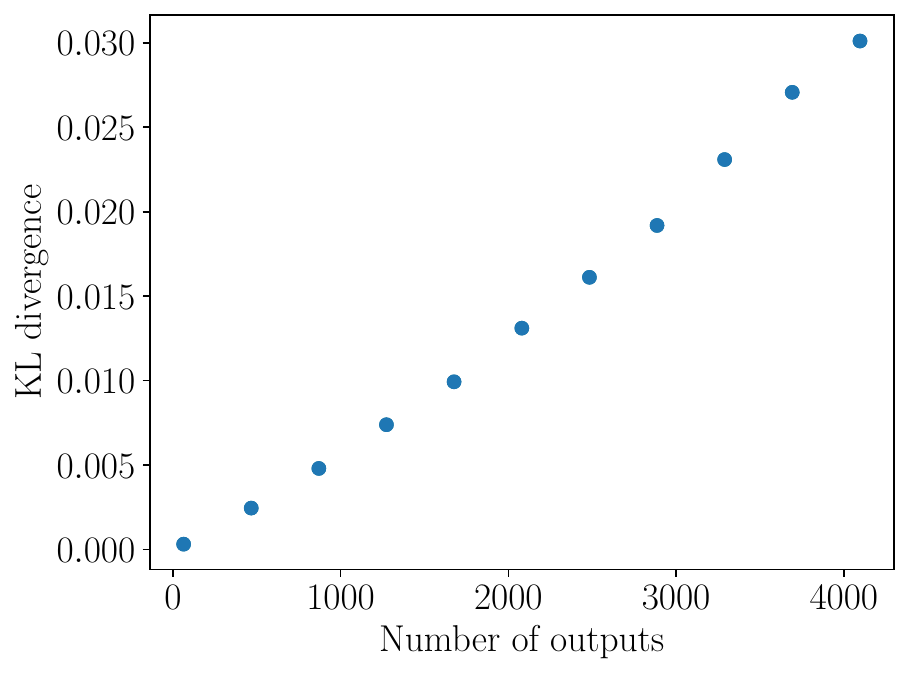"}
	\includegraphics[width=0.45\linewidth]{"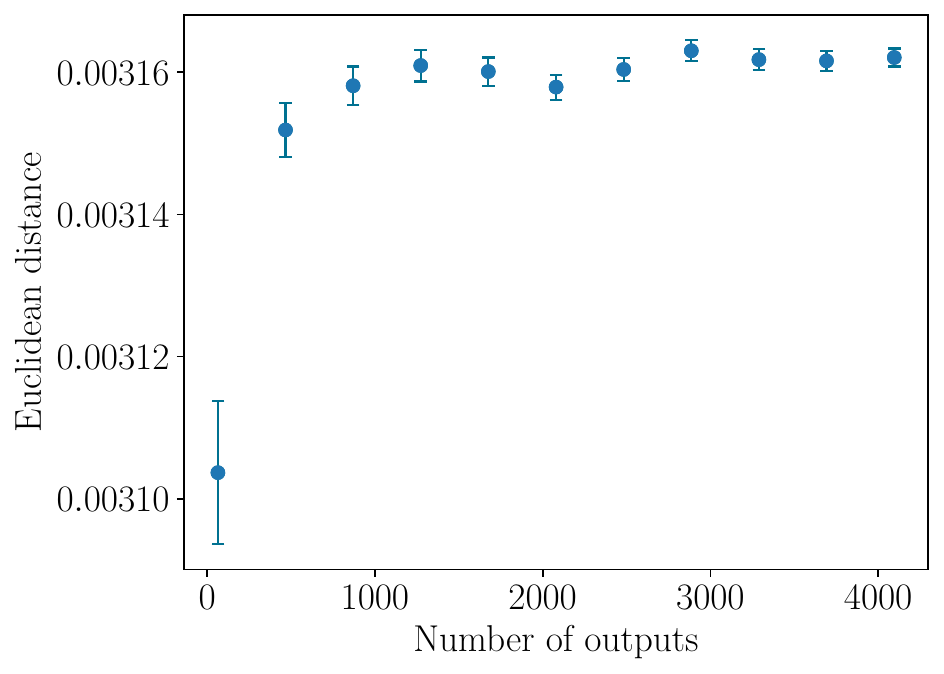"}
	\caption{
        Expected error between a reference distribution, drawn randomly, and $10^3$ distributions obtained from sampling it.
        Results in the left column are for the KL divergence, while those in the right column are for the Euclidean distance.
        The top row shows the dependence on the number of samples for distributions with 64 outcomes, and the bottom row shows the dependence on the number of outcomes, with the number of samples fixed to $10^5$.
        In the bottom-left figure, the error bars are behind the datapoints.
    }
	\label{fig:distr_dist}
\end{figure*}

We sample from a reference, random distribution, and determine the expected error when sampling from that distribution depending on the number of samples, $N_s$.
Figure~\ref{fig:distr_dist} shows the error for different numbers of samples and fixed number of outcomes (top row), and fixed number of samples and different number of outcomes (bottom row).
The KL divergence seems to decrease linearly with the number of samples and also increase linearly with the number of outcomes, and the Euclidean distance appears to decrease as a square root with the number of samples while staying roughly constant with the number of outcomes.
As explained in the main text, these tendencies are in line with known results.
However, in order to give a prescription on the amount of samples to take for a given loss, we need to provide suitable prefactors.
After linear fitting the error to $N_o/N_s$ in the case of the KL divergence, and to $N_s^{-1/2}$ in the case of the Euclidean distance, we arrive at the conclusion that the expressions $N_s=N_o/(2\mathcal{L})$ and $N_s=\mathcal{L}^2$ provide a good guidance for choosing the number of samples for, respectively, the KL divergence and Euclidean distance.

\section{Noise robustness tests on the most promising realizations}
\label{app:robustness}
\setcounter{figure}{0}
In this appendix we plot the distances obtained from the neural network ansatz for the most promising realizations in Fig.~\ref{fig:results_2d_plots}, when we make the sources distribute states mixed with white noise.
This provides further insights into the nonlocality of these distributions since it quantifies how much they can get close to the uniformly random distribution before being likely to admit a local model.
The curves are shown in Fig.~\ref{fig:robustness}.
They suggest noise robustnesses between $10\%$ for the realization in the triangle network and $20\%$ for the realizations in the square and the pentagon with four outcomes.

\begin{figure}[h]
    \centering
    \subfloat[]{
        \includegraphics[width=0.48\linewidth]{"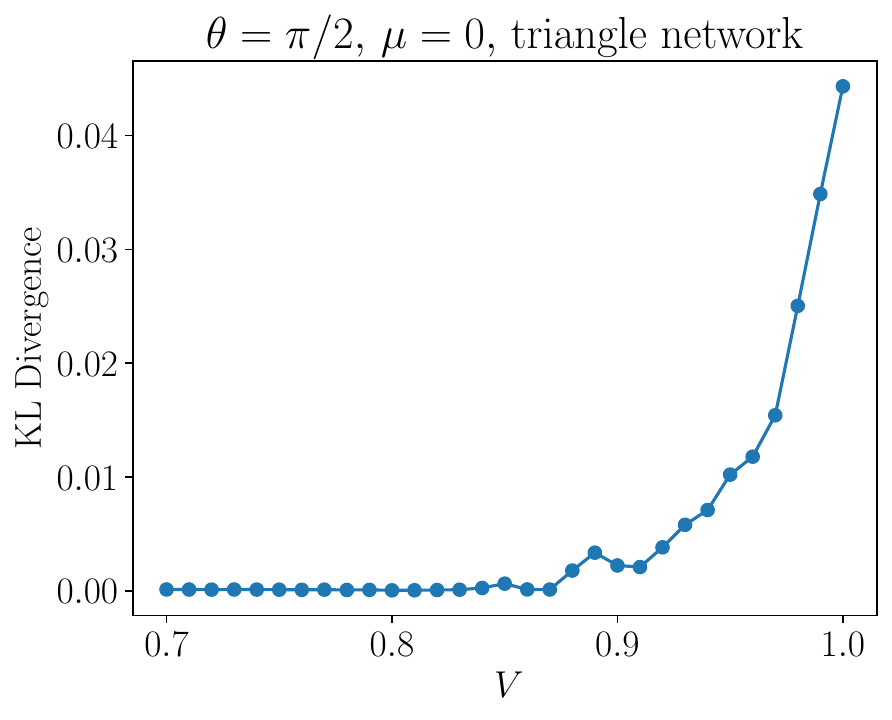"}
    }
    \subfloat[]{
        \includegraphics[width=0.48\linewidth]{"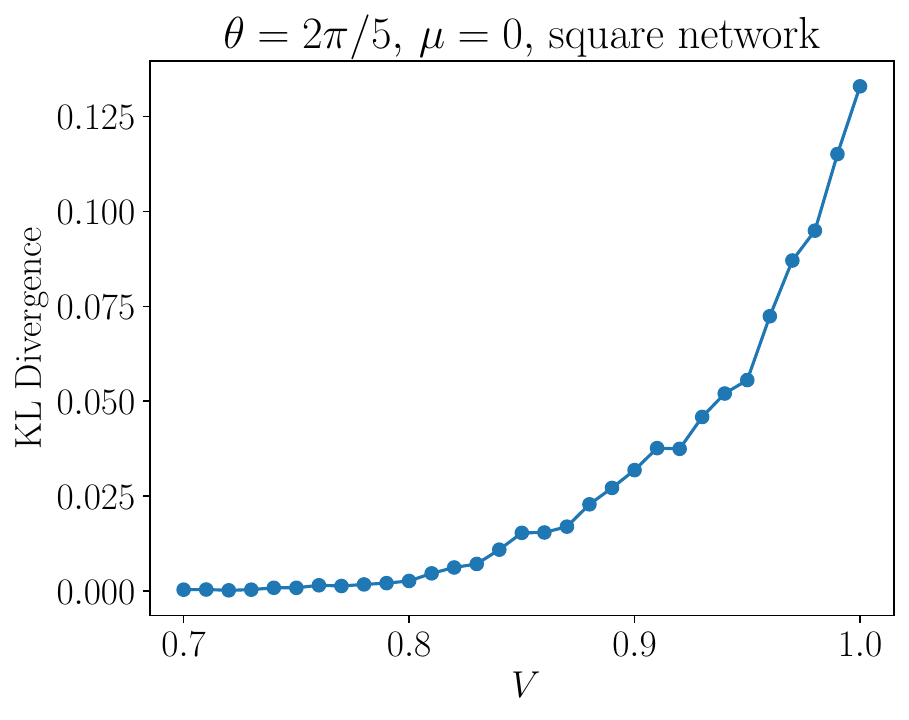"}
    }
    \\
    \subfloat[]{
        \includegraphics[width=0.48\linewidth]{"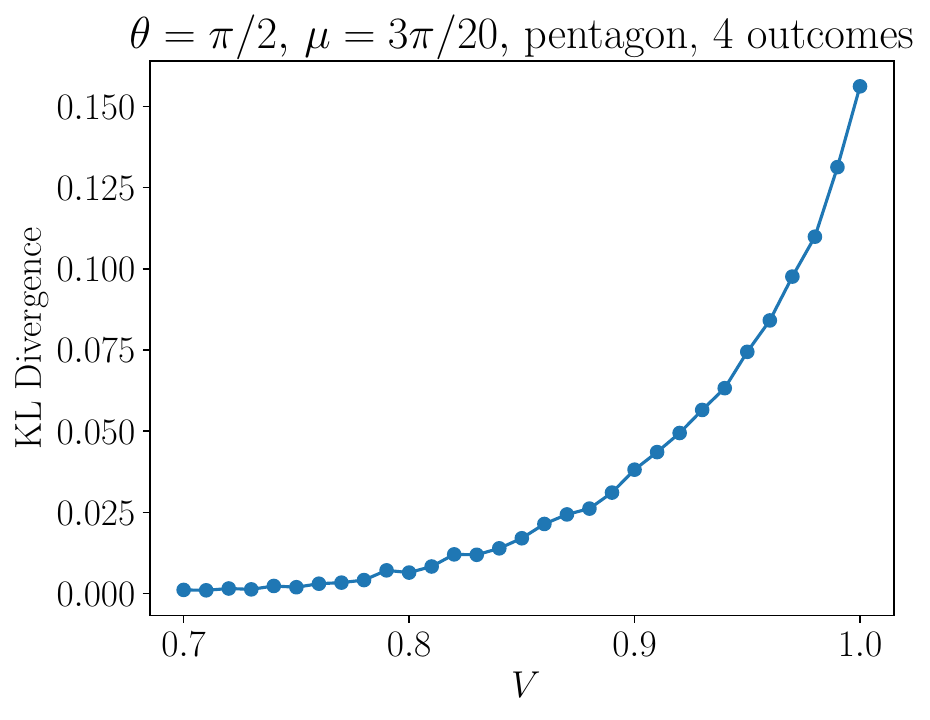"}
    }
    \subfloat[]{
        \includegraphics[width=0.48\linewidth]{"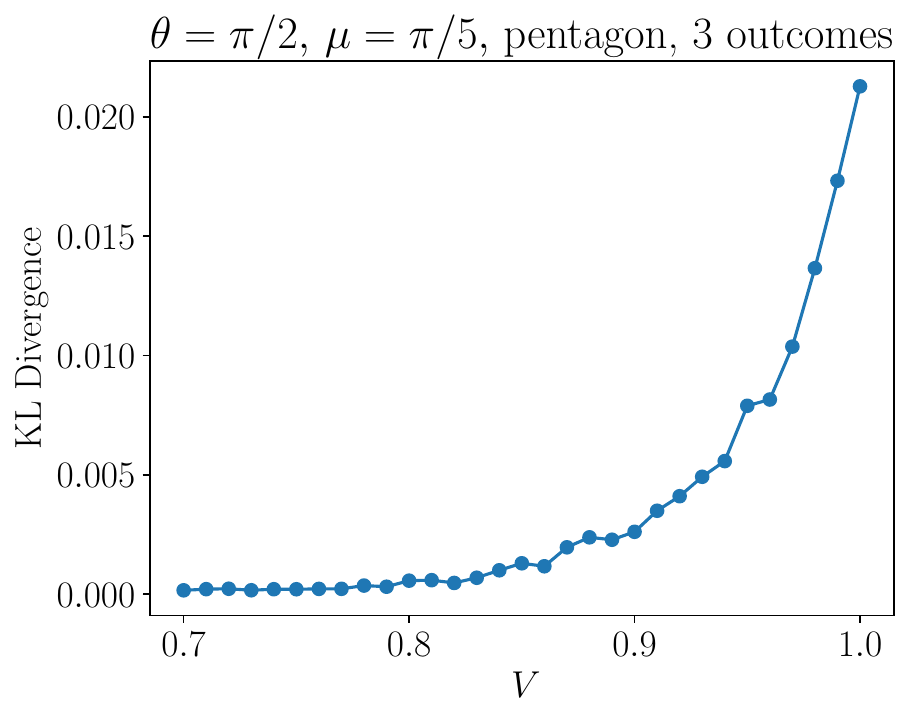"}
    }
    \caption{
        Results of noise robustness of the most promising realizations in (a) the triangle network, (b) the square network, (c) the pentagon network, (d) the pentagon with three outputs per party.
    }
    \label{fig:robustness}
\end{figure}

\setcounter{figure}{0}
\section{More results for ring networks}\label{app:morering}
\begin{figure}[h]
    \centering
    \subfloat[]{
        \includegraphics[width=0.48\linewidth]{"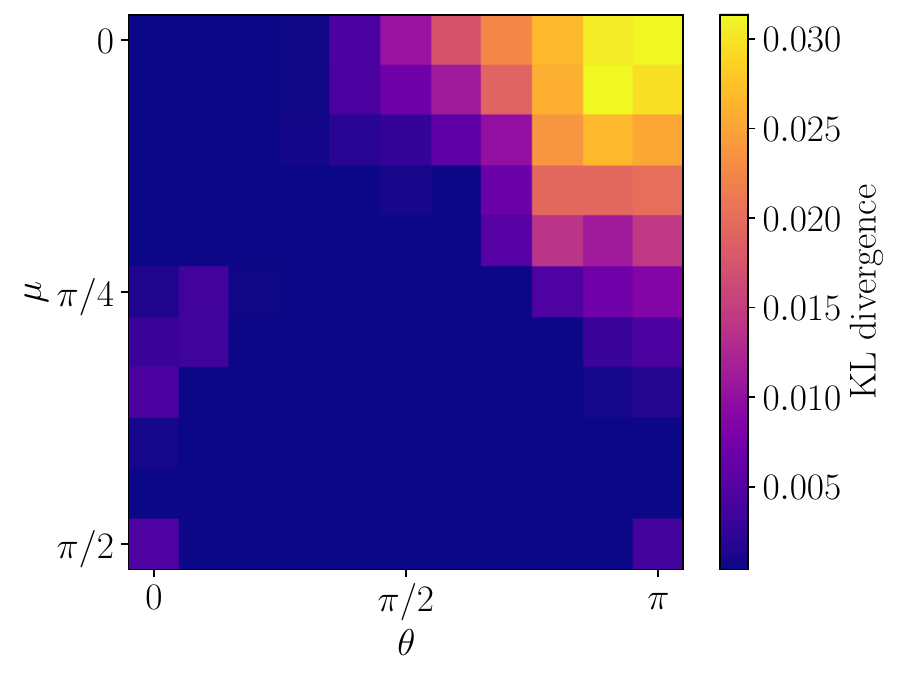"}
        \label{fig:results_2d_plots:triangle2}
    }
    \subfloat[]{
        \includegraphics[width=0.48\linewidth]{"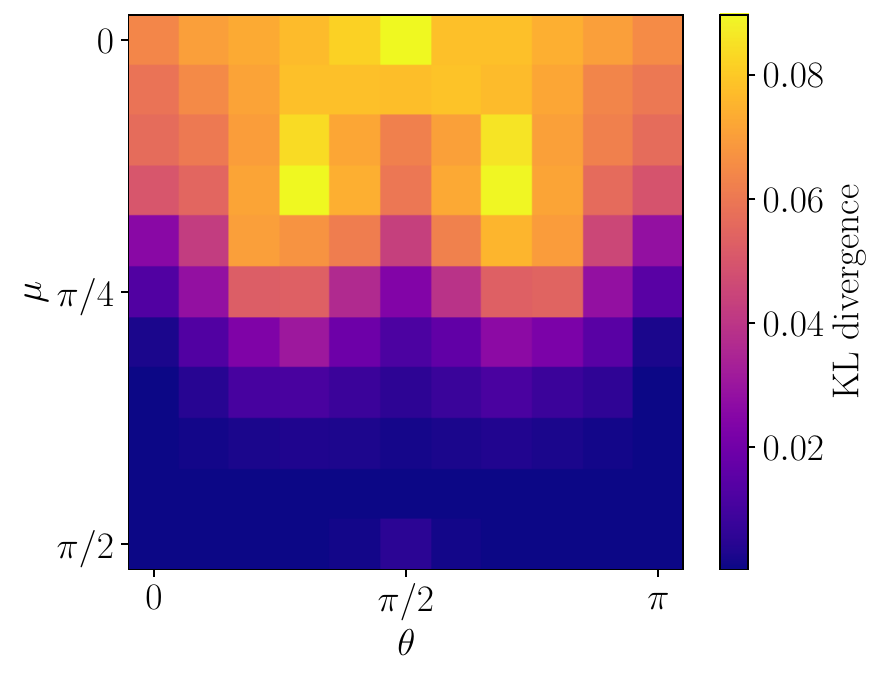"}
        \label{fig:results_2d_plots:square2}
    }
    \\
    \subfloat[]{
        \includegraphics[width=0.48\linewidth]{"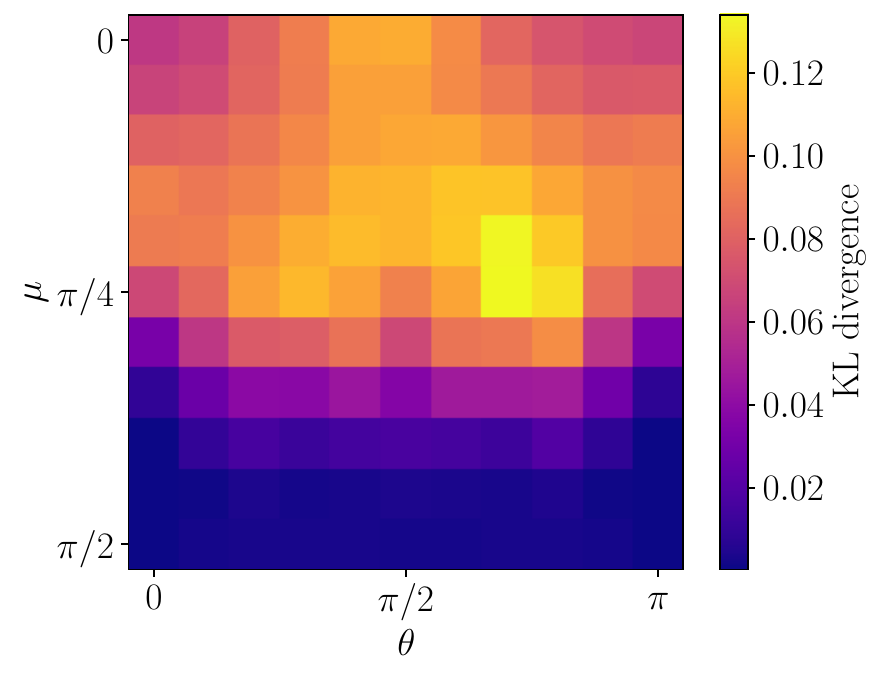"}
        \label{fig:results_2d_plots:pentagon2}
    }
    \subfloat[]{
        \includegraphics[width=0.48\linewidth]{"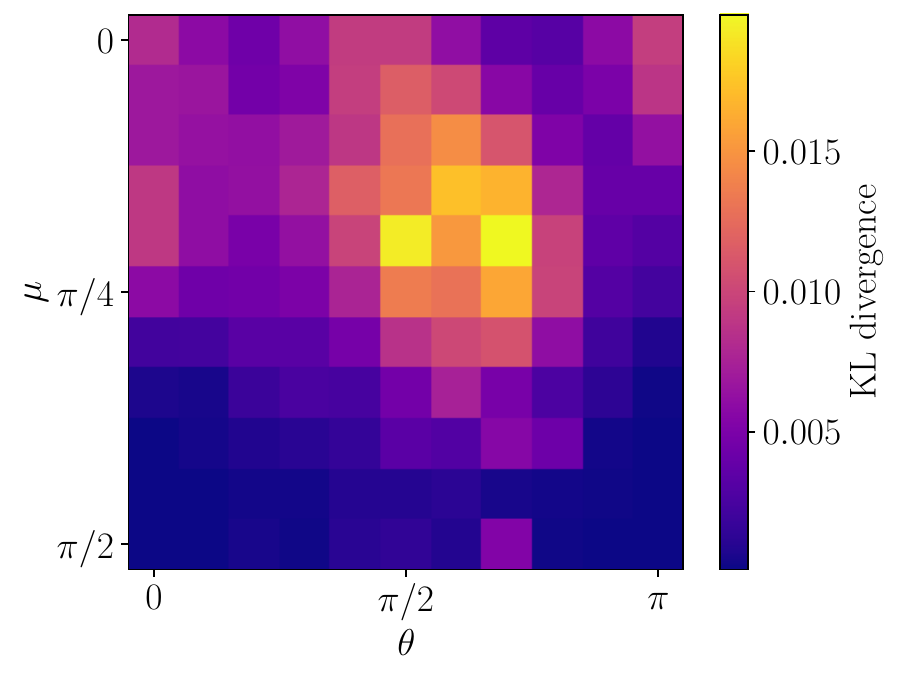"}
        \label{fig:results_2d_plots:pentagon32}
    }
    \caption{Results of the neural network when scanning the family of measurements given in Eq.~\eqref{eq:eBSM} and the family of states in Eq.~\eqref{eq:rotated_state_2} in \protect\subref{fig:results_2d_plots:triangle2} the triangle network, \protect\subref{fig:results_2d_plots:square2} the square network, \protect\subref{fig:results_2d_plots:pentagon2} the pentagon network, and \protect\subref{fig:results_2d_plots:pentagon32} the pentagon with three outputs.
        All results are obtained using a depth of 4 and a width per party of 60.
        Training is performed for a maximum of $10^4$ iterations, stopping earlier if there has been no improvement over $10^3$ iterations.}
    \label{fig:results_2d_plots2}
\end{figure}

In the main text, we have presented results obtained in networks forming loops with states defined in Eq.~(\ref{eq:rotated_state_1}) and measurements defined in Eq.~(\ref{eq:eBSM}).
To complement these, we present a similar analysis for the same family of measurements, but for another family of maximally entangled states given by 
\begin{equation}\label{eq:rotated_state_2}
    \ket{\psi^{\theta}_2}=(e^{\ii\frac{\theta}{2}X}\otimes \id) \ket{\phi^-} = \cos\frac{\theta}{2}\ket{\phi^-} +\ii\sin\frac{\theta}{2}\ket{\psi^-}
\end{equation}
The results are shown in Fig.~\ref{fig:results_2d_plots2} and are indeed quite different from those of Fig.~\ref{fig:results_2d_plots}.
First, we see that the symmetry observed for states around $\theta=\frac{\pi}{2}$ is no longer present in the case of the triangle, but reappears in the larger networks.
In the triangle network, we note that the top-right corner corresponds to the distribution known as the elegant distribution \cite{gisin2017elegantjointquantummeasurement}, which is known to be nonlocal \cite{Gitton2025}.
In the square, the angles describing the most promising states for nonlocality remain the same as in the main text.
In the pentagon the situation changes, the most promising states being far from $\theta=\frac{\pi}{2}$.
However, the most promising measurement still seems to be that for $\mu=\frac{\pi}{4}$.

Reference~\cite{gisin2017elegantjointquantummeasurement} conjectured that the distribution generated by $\ket{\psi^-}$ states and Elegant Joint Measurements was nonlocal in every ring network.
The results in Fig.~\ref{fig:results_2d_plots2} seem to support this conjecture (which for the triangle was confirmed in Ref.~\cite{Gitton2025}), at least for the smaller networks.
However, it is notable that our results seem to indicate that other states and measurements seem to produce distributions further away from the local set.

\end{document}